\documentclass[aps,superscriptaddress,amsmath,amssymb,floatfix,twocolumn,showpacs,amsfonts,longbibliography]{revtex4-1}
\usepackage{times}
\usepackage[varg]{txfonts}
\usepackage{textcomp}
\usepackage{graphicx}
\usepackage{subfigure}
\usepackage{tabu}
\usepackage{color}
\usepackage[colorlinks=true,citecolor=blue,urlcolor=blue,linkcolor=blue,hyperindex]{hyperref}
\usepackage{braket}
\usepackage{float}
\usepackage{overpic}
\usepackage{amssymb}

\allowdisplaybreaks

\begin{document}

\title{Spin excitation spectra of the two dimensional $S=1/2$ Heisenberg model with a checkerboard structure}
\author{Yining Xu}
\affiliation{State Key Laboratory of Optoelectronic Materials and Technologies, School of Physics, Sun Yat-Sen University, Guangzhou 510275, China}

\author{Zijian Xiong}
\affiliation{State Key Laboratory of Optoelectronic Materials and Technologies, School of Physics, Sun Yat-Sen University, Guangzhou 510275, China}

\author{Han-Qing Wu}
\affiliation{State Key Laboratory of Optoelectronic Materials and Technologies, School of Physics, Sun Yat-Sen University, Guangzhou 510275, China}

\author{Dao-Xin Yao}
\email[Corresponding author:]{yaodaox@mail.sysu.edu.cn}
\affiliation{State Key Laboratory of Optoelectronic Materials and Technologies, School of Physics, Sun Yat-Sen University, Guangzhou 510275, China}

\date{\today}

\begin{abstract}
We study the spin excitation spectra of the two-dimensional spin-$1/2$ Heisenberg model with a checkerboard structures using stochastic analytic continuation of the imaginary-time correlation function obtained from a quantum Monte Carlo simulation. The checkerboard models have two different antiferromagnetic nearest-neighbor interactions $J_{1}$ and $J_{2}$, and the tuning parameter $g$ is defined as $J_{2}/J_{1}$. The dynamic spin structure factors are systematically calculated in all phases of the models as well as at the critical points. To give a full understanding of the dynamic spectra, spin wave theory is employed to explain some features of numerical results, especially for the low-energy part. When $g$ is close to $1$, the features of the spin excitation spectra of each checkerboard model are roughly the same as those of the original square lattice antiferromagnetic Heisenberg model, and the high-energy continuum among them is discussed. In contrast to the other checkerboard structures investigated in this paper, the $3\times 3$ checkerboard model has distinctive excitation features, such as a gap between a low-energy gapless branch and a gapped high-energy part that exists when $g$ is small. The gapless branch in this case can be regarded as a spin wave in N$\mathrm{\acute{e}}$el order formed by a "block spin" in each $3\times 3$ plaquette with an effective exchange interaction originating from renormalization. One unexpected finding is that the continuum also appears in this low-energy branch.

\end{abstract}

\maketitle
\date{\today}

\section{Introduction\label{introduction}}
With the evolution of experimental measurement techniques, such as inelastic neutron scattering (INS), and the development of numerical calculation methods, the dynamic signatures of magnetic systems have attracted more attention in recent years. For example, the high-energy portion of the spin excitation spectra of the spin$-1/2$ antiferromagnetic Heisenberg model predicted by spin wave theory deviates from the experimental results~\cite{Christensen, Dalla, ronnow, HeadingsPRL, RIXS}, while the numerical results are well matched with the experiments~\cite{WHZheng2005,sandvik-multimagnon}. Meanwhile, two-magnon excitation spectra were measured in dimerized antiferromagnetic chain material~\cite{chain-experiment}, two-triplon scattering of a cuprate ladder was quantitatively observed \cite{ladder-experiment}, and the structure of the magnetic excitation in a spin$-1/2$ antiferromagnetic triangular lattice Heisenberg system was also studied \cite{triangular-experiment} which is significantly different from the theoretical expectation. Furthermore, dynamical properties also have been studied for some exotic phenomena by using numerical methods, such as the deconfined quantum critical points \cite{deconfined-PRX,Ma-SAC2018} and quantum spin liquids \cite{MZY2018spinliquid}. By studying the excitation spectra of magnetic materials and their related spin models, we can find some new features which can help us to gain a deeper insight of the mechanisms behind these physics.

Numerically, stochastic analytic continuation (SAC) \cite{SAC-Sandvik1998,SAC-Beach2004,SAC-Olav2008,SAC-Thomas2010,Sandvik-SAC2016,Shao-SAC2017} of imaginary-time correlation functions obtained from quantum Monte Carlo (QMC) simulations can be used to study the dynamic spin structure factor $S(q,\omega)$ which can reveal dynamical information about the system. As a method of studying dynamic properties in recent years, the results calculated with the QMC-SAC method have been tested by synthetic data \cite{Shao-SAC2017} and compared with the results calculated with the Bethe ansatz \cite{Sandvik-SAC2016}. Since $S(q,\omega)$ can be accessed directly by INS and nuclear magnetic resonance (NMR) experiments, the most favorable evidence for the validity of the QMC-SAC method is that the calculated $S(q,\omega)$ results are in good agreement with the existing experimental results \cite{Shao-SAC2017}. As an effective numerical method for calculating the complete spectra, the QMC-SAC method has recently been used in some interesting work, such as the spin excitation spectrum of the random singlet state \cite{Shu-SAC}, quantum spin liquids \cite{MZY2018spinliquid}, the dynamical signature of fractionalization at a deconfined quantum critical point \cite{Ma-SAC2018}, and the dynamics of the Higgs mode in spin systems \cite{higgs-SAC,higgs-MZY}. However, unlike the results given by analytical studies, the effects of various modes of spin excitation are intermingled in the results obtained with QMC-SAC. Therefore, the QMC-SAC results should be combined with theoretical explanations to gain further understanding.

Some materials, such as $\mathrm{SrCu_{2}(BO_{3})_{2}}$, have been found to appear like a plaquette phase under certain conditions~\cite{2017SrCu2(BO3)2,2018SrCu2(BO3)2}, in which the spins present a $2\times 2$ periodic block structure. In theoretical studies, such periodic structures can be modeled by checkerboard models (also known as plaquette models in some of the literature). The $2\times 2$ checkerboard model is one of the well-studied spin models. Its ground state properties have been numerically calculated by series expansion \cite{Singh99,Kogaseries}, quamtum Monte Carlo\cite{Wenzelprb,Wenzelprl,CoreOitmaa}, exact diagonalization \cite{Gotze12,Voigt}, real space renormalization groups \cite{RGreview09}, the coupled-cluster method \cite{Gotze12}, and contractor renormalization \cite{CoreOitmaa,CapponiCore}. Some results have also been given through analytical studies like the nonlinear $\sigma$ model \cite{Takano,Kogatheory,Sirkerphonon}, bond operators \cite{Zhitomirsky,Uedaprb,Dorettoprb,Kumartriplon}, and spin wave theory \cite{Kogatheory,Sirkerphonon}.  Interestingly, some similar periodic structures are observed in other experimental materials, such as $\mathrm{Bi}_{2}\mathrm{Sr}_{2}\mathrm{CaCu}_{2}\mathrm{O}_{8+\delta}$ \cite{2002Bi2Sr2CaCu2O8+x,2018Bi2Sr2CaCu2O8+x} and $\mathrm{Ca}_{2-x}\mathrm{Na}_{x}\mathrm{CuO}_{2}\mathrm{Cl}_{2}$ \cite{2004Ca2-xNaxCuO2Cl2}. What is more, the antiferromagnetic clusters with the $3\times 3$ structure are realized in nanomagnets \cite{nanomagnet3-3material,3-3material}. In addition, a feasible method to implement the expected checkerboard models is constructed in the optical lattice~\cite{2002OpticalLattices,2009OpticalLattice,2012OpticalCheckerboard} by cold atom experiments.

\begin{figure}[htbp]
\centering
\includegraphics[width=0.5\textwidth]{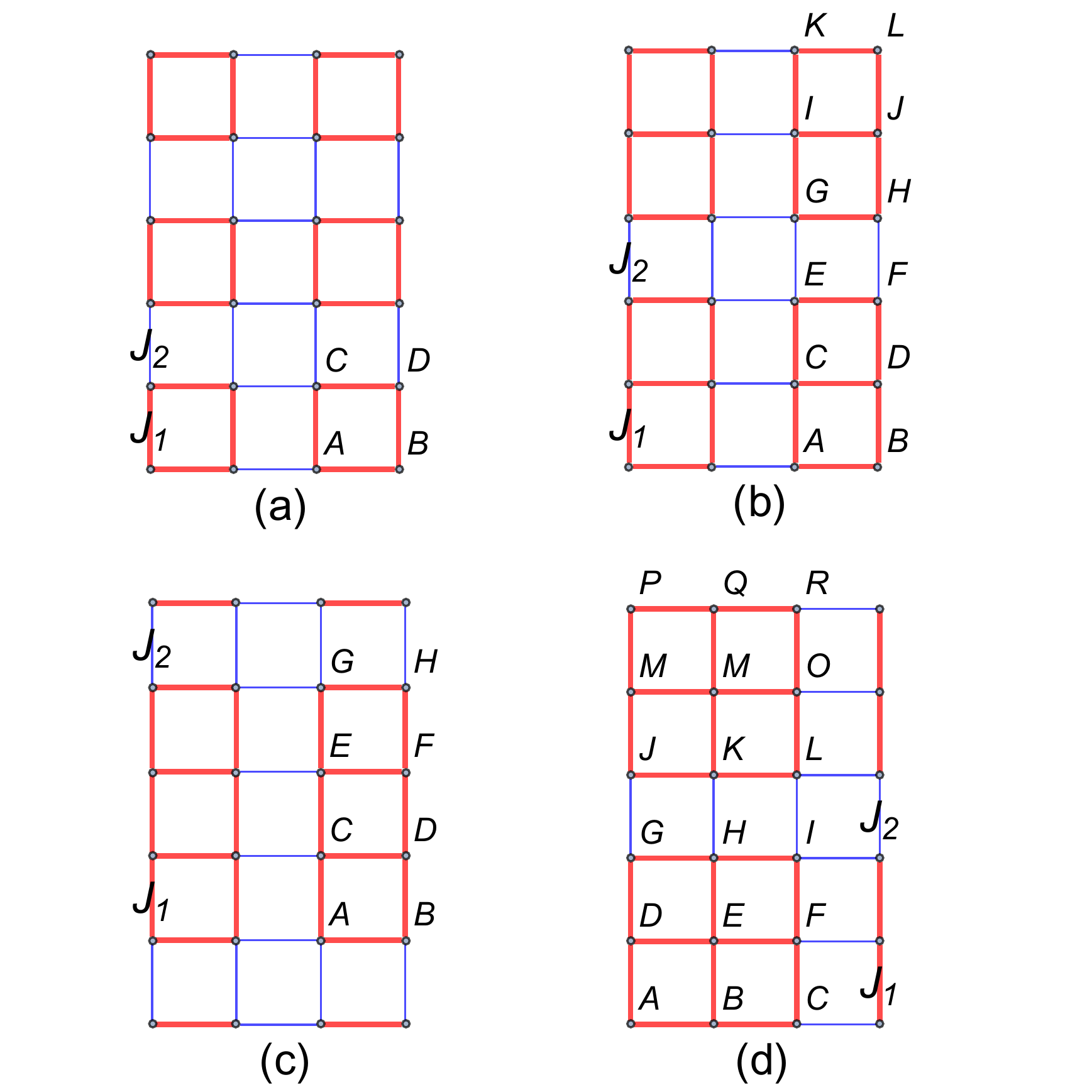}
\caption{\label{model-structure} (Color online) Structures of different checkerboard lattices. (a) $2\times 2$ , (b) $2\times 3$ , (c) $2\times 4$, and (d) $3\times 3$. The intra-sublattice interactions $J_{1}$ and the inter-sublattice interactions $J_{2}$ are represented by thick red and thin blue lines respectively. Here we study the antiferromagnetic case, i.e., $J_{1}>0$, $J_{2}>0$. The tuning parameter $g$ is defined as $g=J_{2}/J_{1}$, where $0<g<1$.  In the paper, we refer to the spin blocks that make up the checkerboard lattice as sublattices.}
\end{figure}
The checkerboard models investigated in this work include the $2\times 2$, $2\times 3$, $2\times 4$, and $3\times 3$ structures. The structures of these checkerboard models are illustrated in FIG.\ref{model-structure}, and we refer to each spin block as a sublattice in this paper. In our recent work \cite{Ran-Ma}, the $O (3)$ universal quantum phase transitions of the $2\times 2$, $2\times 3$, and $2\times 4$ checkerboard models have been studied. The long-range N$\mathrm{\acute{e}}$el order is destroyed at the quantum critical point while the spin-rotation symmetry [$SU(2)$ symmetry] is restored \cite{plaquette-symmetry}. For the disordered phase, i.e., the plaquette phase, the lattice symmetry is not spontaneously broken but is destroyed by the designed model, which is very similar to the so-called dimer phase (or named the coupled-dimer antiferromagnet) \cite{Ma-SAC2018,sachdev2008nphys}.

Although there is some research on the $2\times 2$ checkerboard lattice, the complete dynamical properties in each phase are still lacking. And we are interested in the signatures of the spin excitation spectra with different checkerboard models and the effect of the $O (3)$ quantum phase transition in $S(q,\omega)$. The $3\times 3$ checkerboard model differs from the other three models in that the number of spins in its sublattice is odd and no quantum phase transition is observed by finite-size scaling of conventional physical quantities\cite{Ran-Ma}. We suppose that in this case, the nine spins in a $3\times 3$ sublattice would collectively appear as a "block spin" of spin-$1/2$. For the N$\mathrm{\acute{e}}$el phase of the checkerboard models, $SU(2)$ symmetry is broken simultaneously, and as expected, the Goldstone modes of all the checkerboard models can be described by using spin wave theory. However, the rest of the features that arise as the number of spins in the sublattice increases are equally noteworthy, and we also expect to interpret them from a theoretical perspective.

The rest of the paper is organized as follows: In Sec. \ref{model}, we give the Hamiltonian of the models and briefly introduce the QMC-SAC method. In Sec. \ref{numerical}, we present numerical results for dynamic spin structure factors of different checkerboard models in color plots and describe the features of them. In Sec. \ref{analysis}, the signatures of the low-energy excitation spectra are explained by using spin wave theory, and the high-energy continuum is  discussed in this section. The excitation spectra of the $3\times 3$ model are considered separately due to the model's distinctive features. We summarize our findings in Sec. \ref{conclusion}.

\section{model and numerical method \label{model}}

\subsection{model}
A two-dimensional antiferromagnetic Heisenberg Hamiltonian is considered on checkerboard lattices,
\begin{equation}
H=J_{1}\sum_{\langle i,j\rangle}\mathbf{S}_{i}\cdot \mathbf{S}_{j}+J_{2}\sum_{\langle i,j\rangle'}\mathbf{S}_{i}\cdot \mathbf{S}_{j},
\label{Hamiltonian}
\end{equation}
where $\mathbf{S}_{i}$ denotes the spin-$1/2$ operator on each site $i$; $J_{1}$ and $J_{2}$ are antiferromagnetic nearest-neighbor interactions. The intra-sublattice interactions $J_{1}$ and the inter-sublattice interactions $J_{2}$ correspond to the thick red and the thin blue bonds in FIG.\ref{model-structure}, respectively. The tuning parameter $g$ is defined as $g=J_{2}/J_{1}$, where $g$ takes $0$ to $1$. For the case of $g=1$, no matter which structure these are, they recover an original antiferromagnetic Heisenberg model with uniform interactions. When $g=0$, the interactions between sublattices vanish and the sublattices are isolated from each other.

 In all of these models, the intra-sublattice interaction $J_{1}$ takes a fixed value of $J_{1}=1$, and the inter-sublattice interaction $J_{2}$ varies according to $J_{2}=g\,J_{1}$. 

\subsection{numerical method}
In the process of numerical calculation, the dynamic spin structure factor $S(q,\omega)$ cannot be directly calculated by QMC simulations. In order to obtain $S(q,\omega)$, the imaginary-time correlation function should be measured by using stochastic series expansion (SSE) \cite{Sandvik-SSE,Sandvik-computation}  QMC first.

The imaginary-time correlation function $G_{q}(\tau)$ describes the dynamic spin-spin correlation of a given transferred momentum $q$ in momentum space, which is defined as
\begin{equation}
G_{q}(\tau)=\langle \mathbf{S}_{-q}(\tau) \cdot \mathbf{S}_{q}(0) \rangle .
\end{equation}
Here we consider the isotropic Heisenberg spin, so $G_{q}(\tau)=3\langle S^{z}_{-q}(\tau) S^{z}_{q}(0) \rangle$, where $S^{z}_{q}$ is the Fourier transform of the spin. For a set of imaginary-time points $\{\tau_{i}\}$, the statistical errors of $G_{q}(\tau)$ with the same $q$ obtained from QMC are correlated; therefore the covariance matrix is necessary to express their full characterization. The covariance matrix is given by
\begin{equation}
C_{ij}=\frac{1}{N_b(N_b-1)}\sum_{\alpha=1}^{N_b}(G^{\alpha}(\tau_{i})-\bar{G}(\tau_{i}))(G^{\alpha}(\tau_{j})-\bar{G}(\tau_{j})),
\end{equation}
where $N_b$ is the number of QMC bins and $\bar{G}(\tau_{i})$ is the statistical average of $G^{\alpha}(\tau_{i})$.

From the $G_{q}(\tau)$ for a series of imaginary-time points, $S(q,\omega)$ can be reconstructed using the relation
\begin{equation}
G_{q}(\tau)=\frac{1}{\pi}\int_{-\infty}^{\infty}d\omega S(q,\omega) e^{-\tau \omega}.
\label{S-G}
\end{equation}
To carry out the analytic continuation, we use the SAC \cite{SAC-Sandvik1998,SAC-Beach2004,SAC-Olav2008,SAC-Thomas2010,Sandvik-SAC2016,Shao-SAC2017} approach here, which does not impose explicitly the entropic prior, rather than the maximum entropy method \cite{ME-QMC}. In the SAC process, the spectrum is typically parametrized as the sum of a large number of $\delta$ functions, though other forms have also been proposed \cite{AccelerationME}. A suitable spectrum is sampled in a Monte Carlo simulation using a likelihood function
\begin{equation}
P(S) \propto \exp(-\frac{\chi^{2}}{2\Theta}),
\label{P-SAC}
\end{equation}
where $\Theta$ is a fictitious temperature and $\chi^{2}$ is the goodness of fit. The goodness of fit is defined as
\begin{equation}
\chi^{2}=\sum_{i,j}(G'(\tau_{i})-\bar{G}(\tau_{i}))C_{ij}^{-1}(G'(\tau_{j})-\bar{G}(\tau_{j})),
\label{x2}
\end{equation}
where $G'(\tau_{i})$ is obtained from the current spectrum by using Eq.~(\ref{S-G}). Good, stable results can be judged by $\chi^{2}$. The $\chi^{2}$ is varied with $\Theta$ which acts as a regularization parameter in Eq.~(\ref{P-SAC}). For the choice of $\Theta$, we adopt the temperature-adjustment scheme given in Ref.~\cite{Shao-SAC2017}; the purpose is to adjust $\Theta$ such that $<\chi^{2}>\approx\chi_{min}^{2}+\sqrt{2\chi_{min}^{2}}$. The more detailed technical content can be found in Refs.~\cite{Sandvik-SAC2016,Shao-SAC2017,SM-SAC,Shu-SAC}.

\section{numerical results \label{numerical}}

Here, we extract the $S(q,\omega)$ for the $2\times 2$, $2\times 3$, $2\times 4$, and $3\times 3$ checkerboard models, and display the results in respective color plots, while the checkerboard models are classified into two classes based on whether a finite critical point can be measured by finite-size scaling of conventional physical quantities (like the Binder ratio, uniform magnetic susceptibility and spin stiffness). In order to better present the spin excitation spectra, the color function of the $S(q,\omega)$ shown in this section is a piecewise function, where the low-intensity portion of the result is represented by a linearly distributed color function and the divergent portion (which is less than $1\%$ of the total amount of data) is treated with logarithm.

We use periodic boundary conditions in the two-dimensional $L\times L$ checkerboard lattice with $L=48$ and perform the numerical calculation. The QMC calculations are carried out at inverse temperature $\beta=L$, which gives $T=0$ results \cite{Sandvik-computation,Sandvik-SAC2016} for $G_{q}(\tau)$ at the momentum considered. The calculations are performed in all phases of the models as well as at the critical points.

The SAC method we use here is the standard form that uses a parametrization of the spectral function with a large number of equal-amplitude $\delta$ functions \cite{Shao-SAC2017,Shu-SAC}.
The inverse of the Laplace transform in Eq.(\ref{S-G}) has an ill-posed nature, and the noise in $G_{q}(\tau)$ is one of the key factors to solve this problem. (In principle, the problem can be settled when the errors are small enough.) Therefore, we need the relative statistical errors of $G_{q}(\tau)$ given by QMC $\approx10^{-5}$~\cite{Sandvik-SAC2016}. On the other hand, in order to obtain a reliable result $S(q,\omega)$, the goodness of fit $\chi^{2}$ in Eq.~(\ref{x2}) should be controlled to $\chi^{2}/N_{\tau}\approx1$ \cite{Sandvik-SAC2016}, where $N_{\tau}$ is the number of imaginary-time points we choose, and here we adjust the $N_{\tau}$ to about $50$.

\subsection{checkerboard models with quantum phase transition}
For the models with the $2\times 2$, $2\times 3$, and $2\times 4$ checkerboard structure, the reduction of $J_{2}$ drives the long-range N$\mathrm{\acute{e}}$el order to the plaquette phase through quantum phase transition of the $O (3)$ universality class at $g=g_{c}$, where the $O (3)$ transition is identified by the critical exponent of the finite-size scaling obtained in our recent work\cite{Ran-Ma}. For the $2\times 2$ structure, $g_{c}=0.548524(3)$; for the $2\times 3$ structure, $g_{c}=0.4694(1)$; and for the $2\times 4$ structure, $g_{c}=0.456978(2)$.
\begin{figure*}[htbp]
\centering
\includegraphics[width=0.9\textwidth]{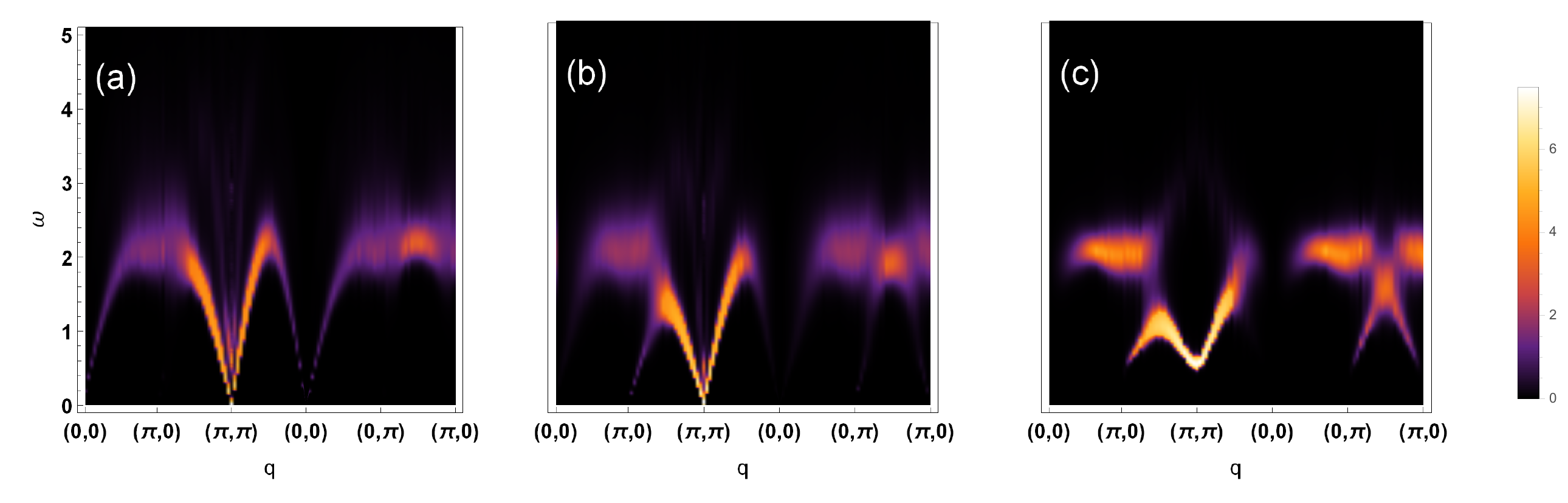}
\caption{\label{QMC-SAC/2-2} (Color online)  Dynamic spin structure factor $S(q,\omega)$ obtained from QMC-SAC calculations for $2\times 2$ checkerboard  model in different $g$, where $g$ is $0.8$ (a), $g_{c} (0.5485)$ (b), and $0.3$ (c). When $g=0.8$, the model is in the N$\mathrm{\acute{e}}$el phase. For $g=0.3$, the model is in the plaquette phase, and the $SU(2)$ symmetry is restored, which appears as gapped spectra.}
\end{figure*}

\begin{figure*}[htbp]
\centering
{\subfigure{
\includegraphics[width=0.9\textwidth]{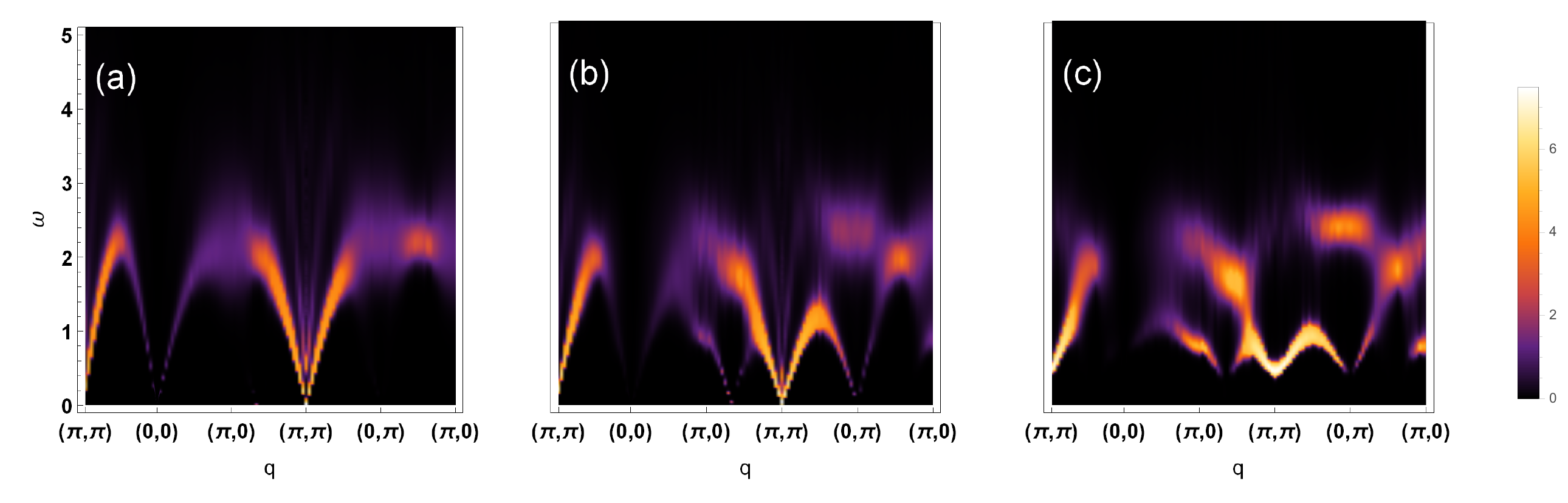}}}
{\subfigure{
\includegraphics[width=0.9\textwidth]{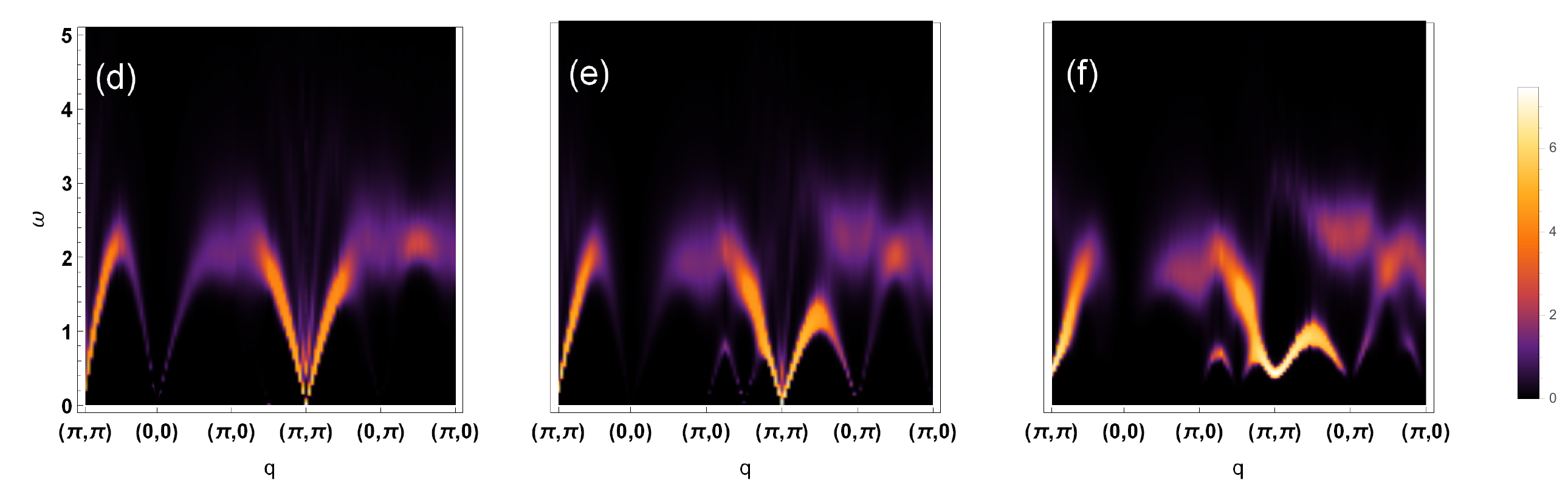}}}
\caption{\label{QMC-SAC/3&4} (Color online) Dynamic spin structure factor $S(q,\omega)$ obtained from QMC-SAC calculations for the $2\times 3$ [(a)--(c)] and $2\times 4$ [(d)--(f)] checkerboard models with different $g$. When $g\neq 1$, the symmetry changes from the original $C_{4}$ symmetry to the $C_{2}$ symmetry. The spectra of the N$\mathrm{\acute{e}}$el phase are shown in (a) and (d), where $g$ is $0.75$ and $0.7$ for the $2\times 3$ and $2\times 4$ models, respectively. In (b) and (e), $g$ takes a critical value $g_{c}$; $g_c=0.4694$ for $2\times 3$, and $g_c=0.4569$ for $2\times 4$. The gapped spectra of the plaquette phase with $g=0.25$ are shown in (c) and (f).}
\end{figure*}

\begin{figure}[htbp]
\centering
{\subfigure[]{
\includegraphics[width=0.21\textwidth]{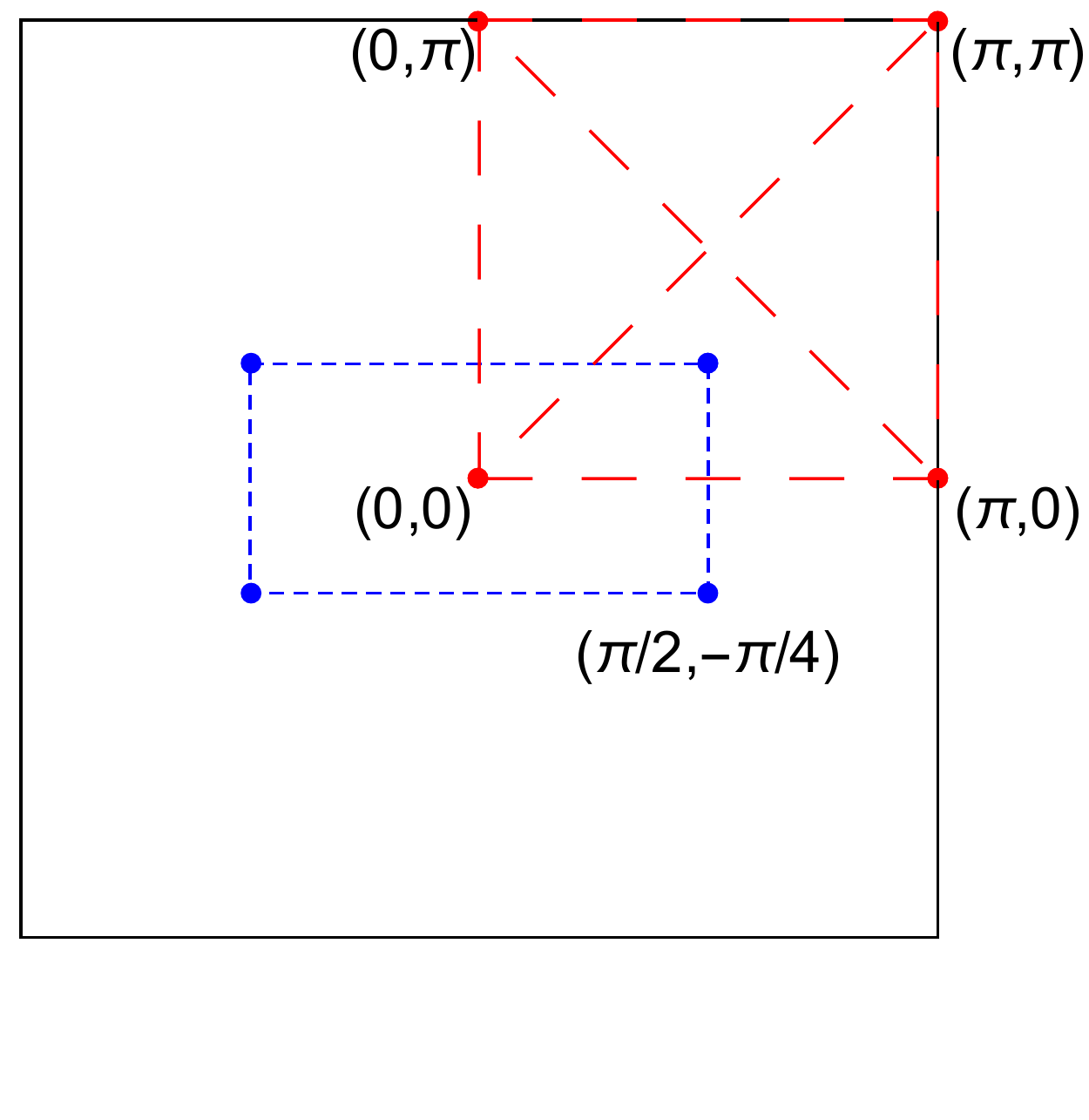}}}
{\subfigure[]{
\includegraphics[width=0.25\textwidth]{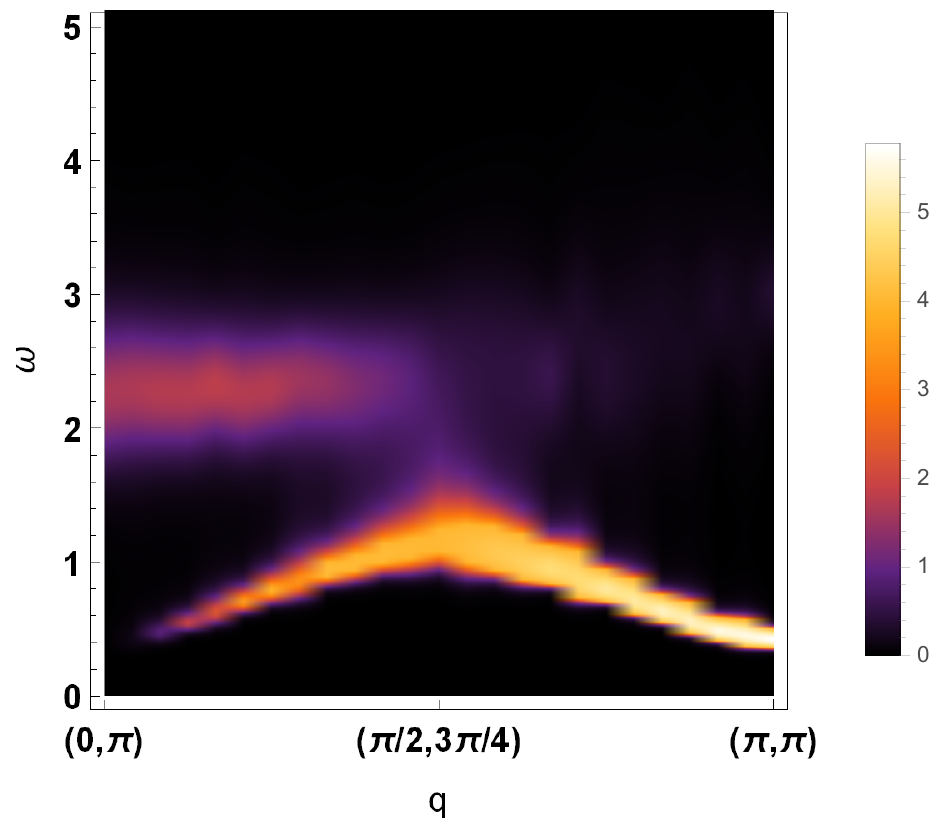}}}
\caption{\label{BZ} (Color online) (a) The blue dashed lines mark the folded Brillouin zone of the $2\times 4$ checkerboard model and the red dashed lines represent the wave vector path we chose. (b) Dynamic spin structure factor obtained from QMC-SAC calculations on the $(0,\pi) \rightarrow (\pi/2,3\pi/4) \rightarrow (\pi,\pi)$ path for the $2\times 4$ checkerboard  model with $g=0.25$. The low-energy modes appear at $(\pi,\pi)$ and $(0,\pi)$, and the low-energy part of the spectrum has a periodic structure.}
\end{figure}

The $S(q,\omega)$ results of the $2\times 2$ checkerboard model are shown in FIG.\ref{QMC-SAC/2-2} along the path $(0,0) \rightarrow (\pi,0) \rightarrow (\pi,\pi) \rightarrow (0,0) \rightarrow (0,\pi) \rightarrow (\pi,0)$, and we analyze the salient features of the spin excitation spectra. In the N$\mathrm{\acute{e}}$el phase, the gapless Goldstone mode is observed at $(\pi,\pi)$ as shown in FIG.\ref{QMC-SAC/2-2}(a), where the spectral weight is well known to be divergent. And the gapless behavior appears at $(0,0)$ also, but the spectral weight tends to vanish approaching $(0,0)$ as expected due to the conservation of total $S^{z}$. For the case of $g=g_{c}$ in FIG.\ref{QMC-SAC/2-2}(b), in addition to the two gapless points as in the N$\mathrm{\acute{e}}$el phase, gapless excitation also occurs at $(\pi,0)$ and $(0,\pi)$, and they are ascribed to the effect of Brillouin zone folding. One noteworthy feature is that the "V" shaped structure around $(\pi,\pi)$ extends to very high energy when $g\geq g_c$, and it also appears in the other 3 checkerboard structures and even in the easy-plane $J_1-J_2$ model \cite{Ma-SAC2018}.
When $g=0.3$, the spin system loses magnetic order due to the formation of spin singlets and is dominated by the disordered ground state with $SU(2)$ symmetry. As shown in FIG.\ref{QMC-SAC/2-2}(c), all spin excitations are gapped; the spectral weight of the high-energy excitation partially increases and the widths of the continua become narrower, which implies that the spin system no longer has a magnetic order; and as $g$ decreases, the excitations between the sublattices tend to disappear and thus the spin excitations should be concentrated primarily in the sublattice.

For the models with the $2\times 3$ and $2\times 4$ checkerboard structure, the results of $S(q,\omega)$ are shown in FIG.\ref{QMC-SAC/3&4}. For these two models, in order to reflect the asymmetry of interaction between $x$ and $y$ directions in the spin excitation spectra, we choose the path $(\pi,\pi) \rightarrow (0,0) \rightarrow (\pi,0) \rightarrow (\pi,\pi) \rightarrow (0,\pi) \rightarrow (\pi,0)$. When $g$ is no longer equal to $1$, the ground state of these models is still in the N$\mathrm{\acute{e}}$el phase and the features of $S(q,\omega)$ in FIG.\ref{QMC-SAC/3&4}(a) and (d) are not significantly different from the results of the $2\times 2$ checkerboard model, although the symmetry of the interactions changes from $C_{4}$ to $C_{2}$. But when $g=g_{c}$, the difference becomes apparent. Along the path $(\pi,0)\to (\pi,\pi)$ in FIG.\ref{QMC-SAC/3&4}(b) and (e), we can see that for the $2\times 3$ model, there is a gapless excitation at $(\pi, \pi/3)$, and for the $2\times 4$ model, gapless excitation appears at $(\pi, 0)$ and $(\pi, \pi/2)$; these are also from the effect of Brillouin zone folding. And in this path, the low-energy part of spin excitation spectra exhibits a folding feature. In addition, gapless excitations also appear at $(0,\pi)$ in FIG.\ref{QMC-SAC/3&4}(b) and (e). As $g$ decreases, the ground state of the spin system becomes the plaquette phase, and all spin excitations are gapped, but the periodic structures of $S(q,\omega)$ survive, as shown in FIG.\ref{QMC-SAC/3&4}(c) and (f).

The periodic interactions of the designed model lead to the effect of Brillouin zone folding whether in the phase with a long-range order or at the critical point \cite{Kogatheory,Sirkerphonon}. This is illustrated by the spectrum in FIG.\ref{QMC-SAC/2-2}(b), FIG.\ref{QMC-SAC/3&4}(b) and (e), where the low-energy modes not only appear at wave vectors $(\pi,\pi)$ and $(0,0)$ but also at other wave vectors along the path $(\pi,0) \rightarrow (\pi,\pi) \rightarrow (0,\pi)$ due to the folded $(\pi,\pi)$. $S(q,\omega)$ at $(0,0)$ is folded also but the low-energy mode can hardly be observed along the path $(0,\pi) \rightarrow (0,0) \rightarrow (\pi,0)$ as a result of the negligible spectral weight around the Brillouin zone center. From the spin excitation spectra shown in FIG.\ref{QMC-SAC/2-2}(c), FIG.\ref{QMC-SAC/3&4}(c) and (f), we can find that the wave vectors of low-energy modes are the same as the corresponding critical spectra of each structure, and therefore whether this means that the effect of Brillouin zone folding is also present in the plaquette phase. To answer this question, we should know the excitation inside the Brillouin zone, not just along the boundary. We calculate the $S(q,\omega)$ on the path $(0,\pi) \rightarrow (\pi/2,3\pi/4) \rightarrow (\pi,\pi)$ for the $2\times 4$ checkerboard  model with $g=0.25$ and present it in FIG.\ref{BZ}, which shows the low-energy modes at $(\pi,\pi)$ and $(0,\pi)$. Thus, we can believe that although the plaquette phase is magnetically disordered, the effect of Brillouin zone folding still exists.

\subsection{$3\times 3$ checkerboard model}
\begin{figure*}[htbp]
\centering
\includegraphics[width=0.9\textwidth]{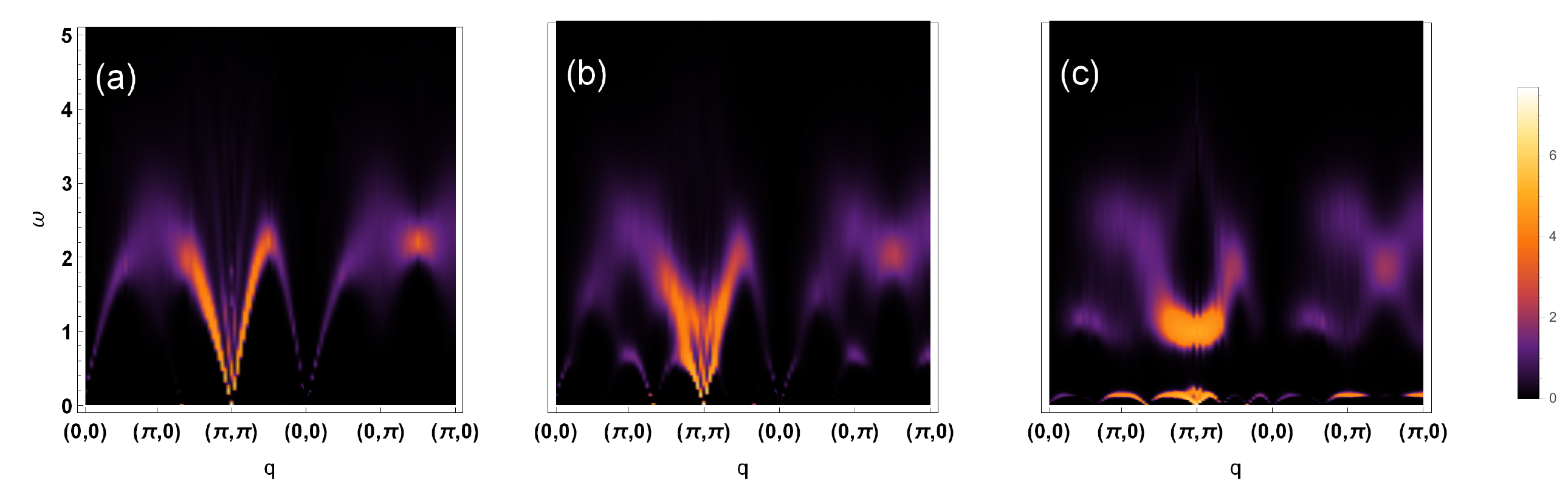}
\caption{\label{QMC-SAC/3-3} (Color online) Dynamic spin structure factor $S(q,\omega)$ obtained from QMC-SAC calculations for the $3\times 3$ checkerboard  model with different tuning parameters $g$, where $g$  is taken as $0.7$ (a), $0.4$ (b), and $0.1$ (c). From these results, we can see that no matter how $g$ changes, the gapless mode at $(\pi,\pi)$ persists. The spectrum of $g=0.1$ is different from the spectra of other structures; especially, the gapless branch is completely separated from the gapped part.
}
\end{figure*}

The $3\times 3$ checkerboard model is very special, with $9$ spins in each sublattice, which means that one spin is likely to be unpaired. When $g$ changes, there is no quantum phase transition observed. We suspect this is due to the existence of unpaired spins in each sublattice; more precisely, each $3\times 3$ sublattice is "renormalized" to an effective "block spin" with spin-$1/2$. As discussed in the previous section, at the limits of $g = 1$ and $g = 0$, the ground states of the spin system are clearly different. In order to know how the change occurs when $g$ decreases, we calculate the $S(q,\omega)$ by taking $g=0.7, 0.4$, and $0.1$ to discuss the $3\times 3$ checkerboard model from the perspective of dynamic signatures. The numerical results are shown in FIG.\ref{QMC-SAC/3-3}, in which wave vectors takes the path of $(0,0) \rightarrow (\pi,0) \rightarrow (\pi,\pi) \rightarrow (0,0) \rightarrow (0,\pi) \rightarrow (\pi,0)$.

When $g=0.7$, $S(q,\omega)$ as shown in FIG.\ref{QMC-SAC/3-3}(a) have the same basic features in the spin excitation spectra as the N$\mathrm{\acute{e}}$el phase obviously. When $g=0.4$, the low-energy portion of the $S(q,\omega)$ shown in FIG.\ref{QMC-SAC/3-3}(b) exhibits a periodic structure, which is similar to the $S(q,\omega)$ of the $2\times 3$ model. Especially, the result along the path $(0,0) \rightarrow (\pi,0) \rightarrow (\pi,\pi)$ is almost the same as the corresponding result in FIG.\ref{QMC-SAC/3&4}(b). But as $g$ keep decreasing, when $g=0.1$, we find that the spin excitation spectra become completely different from the previous results.

As shown in FIG.\ref{QMC-SAC/3-3}(c), when $g=0.1$, the gapless branch in the spectrum is totally separated from the high-energy part. This can be seen from the fact that the gapless branch has energy around $J_{2}$, which should be the excitation among the sublattices. And the existence of this gapless branch with a periodic structure proves our "block spin" guess. Since each $3\times 3$ sublattice is "renormalized" into a "block spin", we can regard it as an effective spin system with elongated lattice constant, and the Brillouin zone is reduced by three times in both the $x$ and $y$ directions correspondingly, so the results are repeated three times on all paths. Moreover, since the gapless branch has the obvious character of the spin excitation spectrum in the N$\mathrm{\acute{e}}$el order, we can infer that the effective spin system has long-range N$\mathrm{\acute{e}}$el order.

The high-energy gapped part has energy around $J_{1}$, which should be the excitation inside the sublattice. Unlike the $S(q,\omega)$ of $g=0.4$ [FIG.\ref{QMC-SAC/3-3}(b)], the high-energy part intersects the gapless branch, and they all have a gapless excitation at $(\pi, \pi )$. When $g=0.1$, there is a significant energy gap between the gapless branch and the gapped part, so it is obvious that the spin configuration within the sublattice must not be N$\mathrm{\acute{e}}$el order but is disordered to some extent. Therefore, we suspect that when $g$ is small enough, the ground state of the $3\times 3$ checkerboard model may also have disordered valence bonds similar as to a checkerboard model with finite $g_c$. Finally, the question can be raised as to whether a different physical quantity can be defined to describe this kind of case, which simultaneously exhibits long-range ordered and disordered features.

\section{Analysis\label{analysis}}
In this section, we provide some account for the spectral features of the numerical spectra. From the previously shown color plots, some common features in our numerical spectra can be identified:

(a) When $g$ is large the overall shapes of spectra of the checkerboard models [shown in Fig.\ref{QMC-SAC/2-2}(a) ,Fig.\ref{QMC-SAC/3&4}(a) and (d), Fig.\ref{QMC-SAC/3-3}(a)] are almost the same as the well-known results in an antiferromagnetic square lattice with N$\mathrm{\acute{e}}$el order (which is referred to as the bipartite case hereafter comparing with an enlarged magnetic unit cell in a checkerboard lattice). To be specific, the spectra behave as varying spectral weight along the dispersion curve $\epsilon_{\textbf{k}} \propto \sqrt{1-\gamma_{\textbf{k}}^{2}}$ with $\gamma_{\textbf{k}}=(\cos{k_x}+\cos{k_y})/2$, and the spectral weight diverges at the $(\pi,\pi)$ due to the N$\mathrm{\acute{e}}$el order. Another noteworthy feature is the prominent high-energy continuum in all structures, and it is discussed in Sec. \ref{HEC}.

(b) As $g$ decreases, some low-energy branches appear, which can be seen in Fig.\ref{QMC-SAC/2-2}(b) ,Figs.\ref{QMC-SAC/3&4}(b) and (e) and Fig.\ref{QMC-SAC/3-3}(b). To have a qualitative understanding of these low-energy branches, the dispersions and transverse dynamic structure factors calculated  by using linear spin wave theory are shown in Sec. \ref{LSWT}.

(c) When $g < g_c$, the spectra of the $2\times 2$, $2\times 3$, and $2\times 4$ checkerboard models are gapped [as shown in Fig.\ref{QMC-SAC/2-2}(c) and Figs.\ref{QMC-SAC/3&4}(c) and 3(f)], but the overall shape does not change a lot. However, the spectra of the $3\times 3$ structure are gapless in all taken $g$ owing to the persistent long-range order, and they are discussed in Sec. \ref{3-3model}.

\subsection{Low-energy excitation\label{LSWT}}
To qualitatively study the low-energy branches, we use the linear spin wave theory (LSWT) outlined in the Appendix.\ref{appendix} to calculate the zero-temperature dynamic structure factor of the $2\times 2$, $2\times 3$, and $2\times 4$ checkerboard models. It is well known that standard spin wave theory starts with the assumption of a magnetic ordered phase and may not give the correct prediction about the critical behavior, so we use spin wave theory to study the ordered phase with $g$ close to $g_c$ and try to understand the spectra.

\subsubsection{$g$ close to $g_c$}
\begin{figure}
\flushleft
{\subfigure[]{
\includegraphics[width=0.22\textwidth]{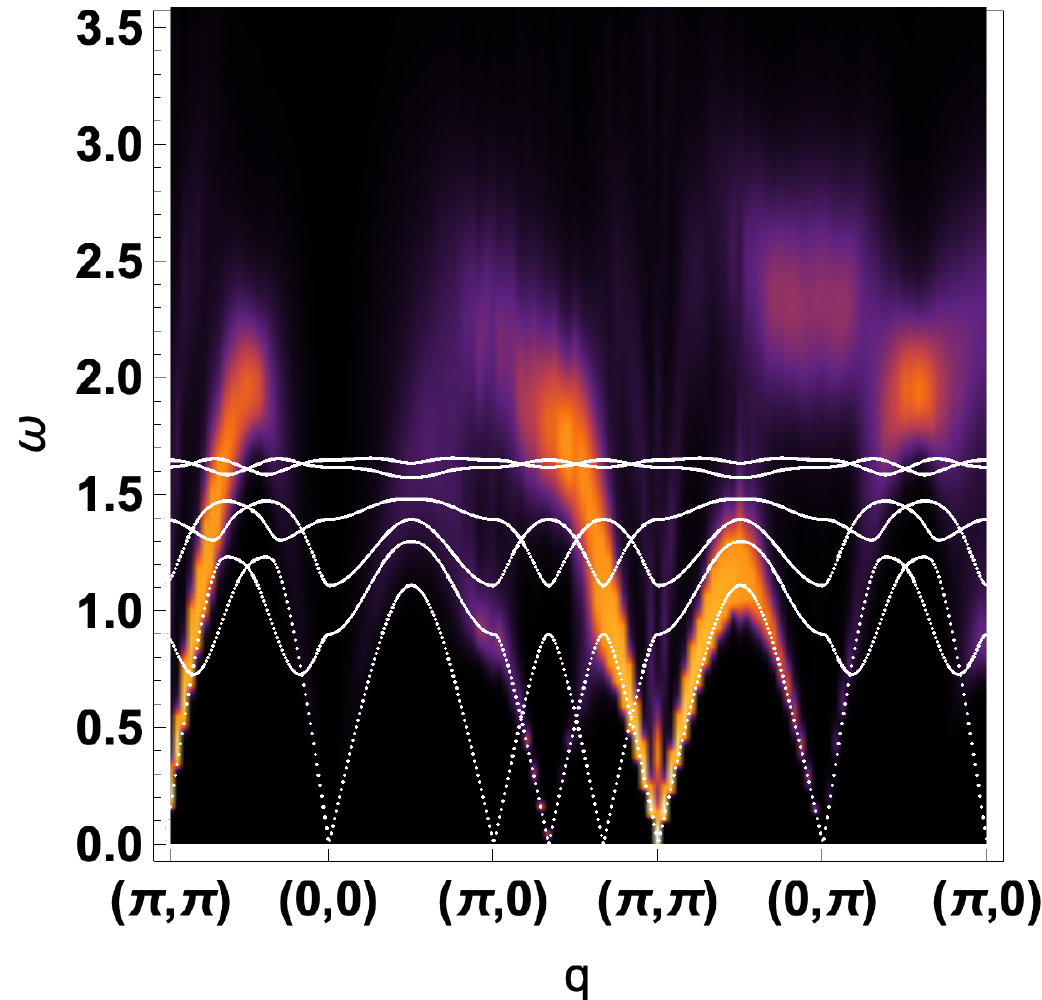}}}
{\subfigure[]{
\includegraphics[width=0.22\textwidth]{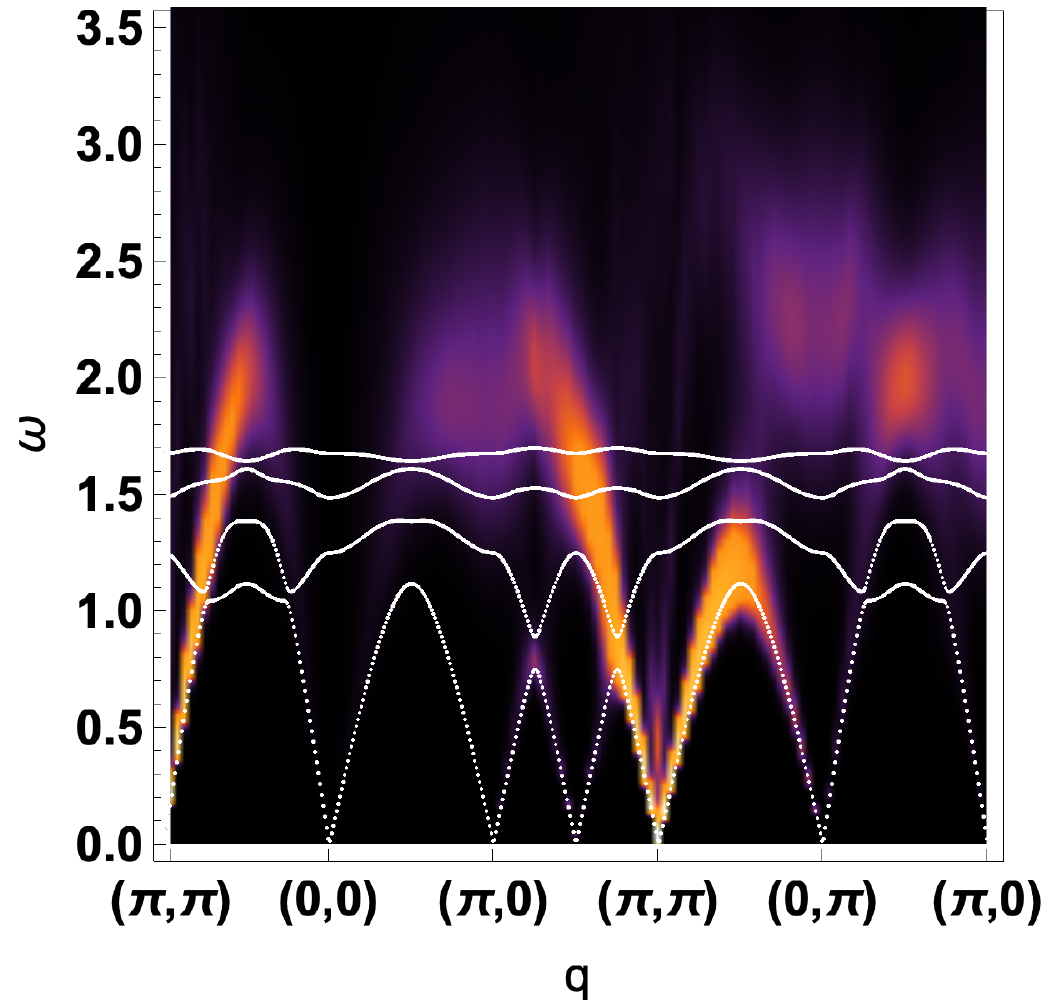}}}

{\subfigure[]{
\includegraphics[width=0.23\textwidth]{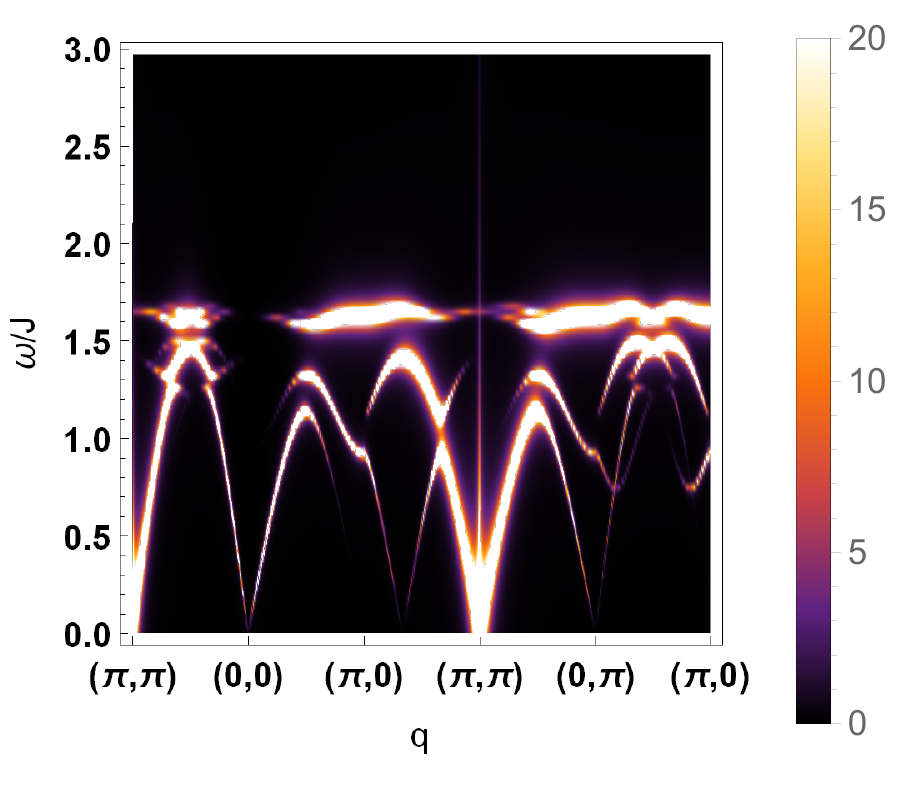}}}
{\subfigure[]{
\includegraphics[width=0.23\textwidth]{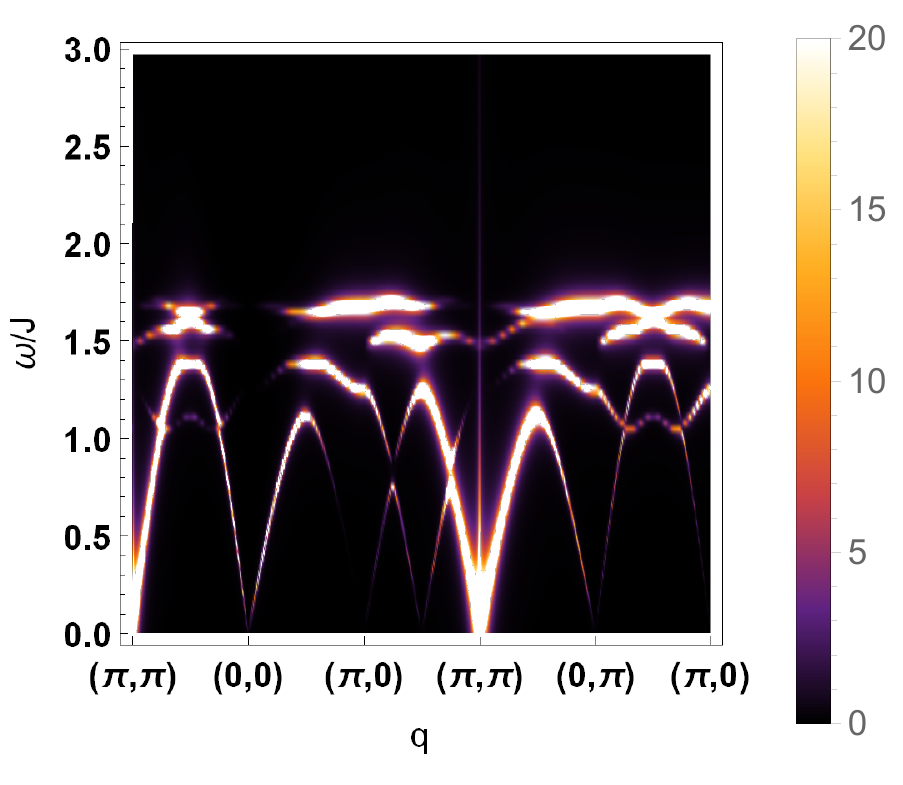}}}
\caption{\label{3&4swt} (Color online) The linear spin wave results of the $2\times 3$ checkerboard model with $g=0.5$ are shown in (a) and (c), and the results of $2\times 4$ with $g=0.47$ are shown in (b) and (d). Here $g$ takes a value close to the critical point to ensure the system is in the ordered phase. In (a) and (b), the linear spin wave dispersions are represented by white dashed line with the color plots in Figs.\ref{QMC-SAC/3&4} (b) and 3(e) as background, and (c) and (d) are the corresponding zero-temperature dynamic structure factors.}
\end{figure}

When $g$ is close to $g_c$, dynamic structure factors (DSFs) and dispersions in the linear spin wave level of the models with $2\times 2$, $2\times 3$, and $2\times 4$ checkerboard structures are presented in Figs.\ref{2-2swt}(a) and (b) and FIG.\ref{3&4swt}. Comparing them with the corresponding numerical spectra in Fig.\ref{QMC-SAC/2-2} (b) and Figs.\ref{QMC-SAC/3&4} (b) and (e), we find that the shape of low-energy dispersions in LSWT matches quite well with the numerical spectra. According to the definition of a DSF (which consists of the matrix element of the spin operator and a $\delta$ function; see Appendix. \ref{appendix}), these matchings mean the low-energy dispersions of these systems are described by a spin wave. Concretely speaking:

 \begin{figure}
\centering
{\subfigure[]{
\includegraphics[width=0.22\textwidth]{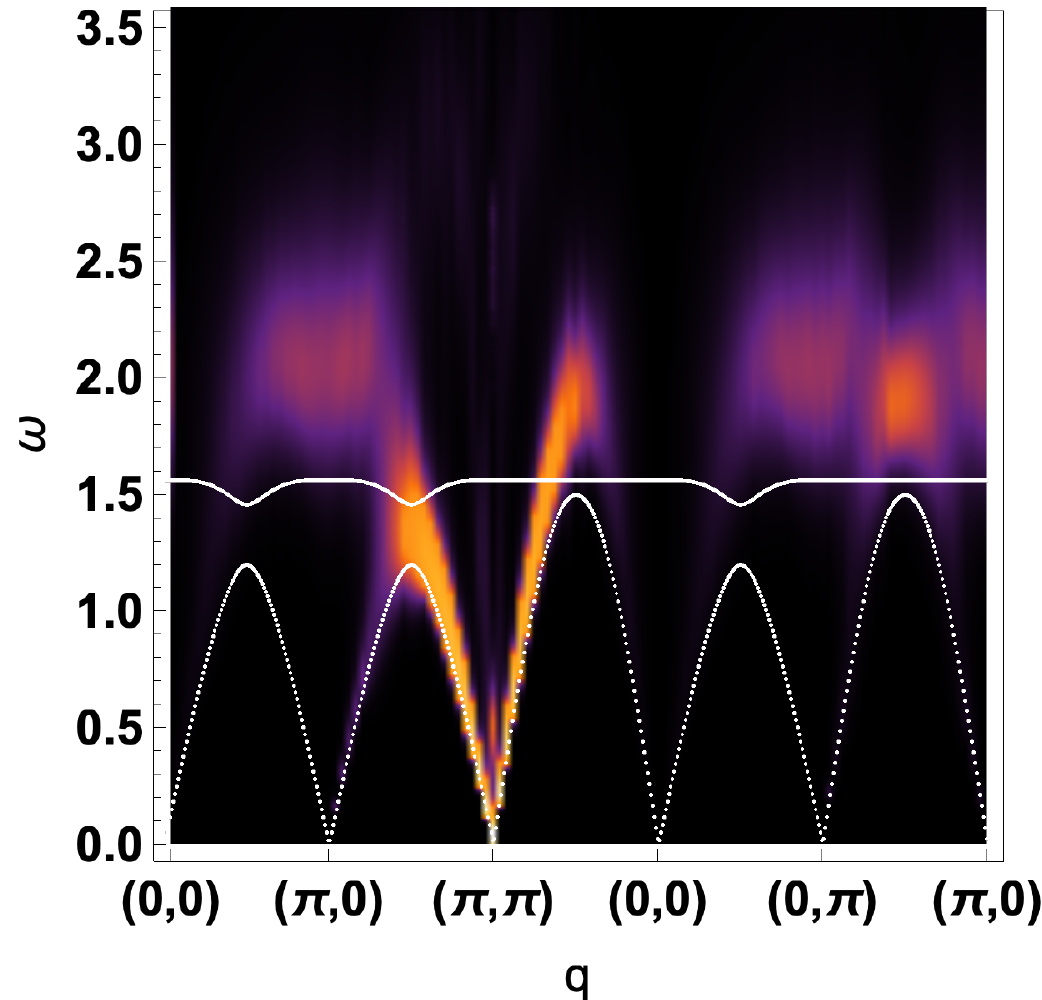}}}
\flushleft
{\subfigure[]{
\includegraphics[width=0.23\textwidth]{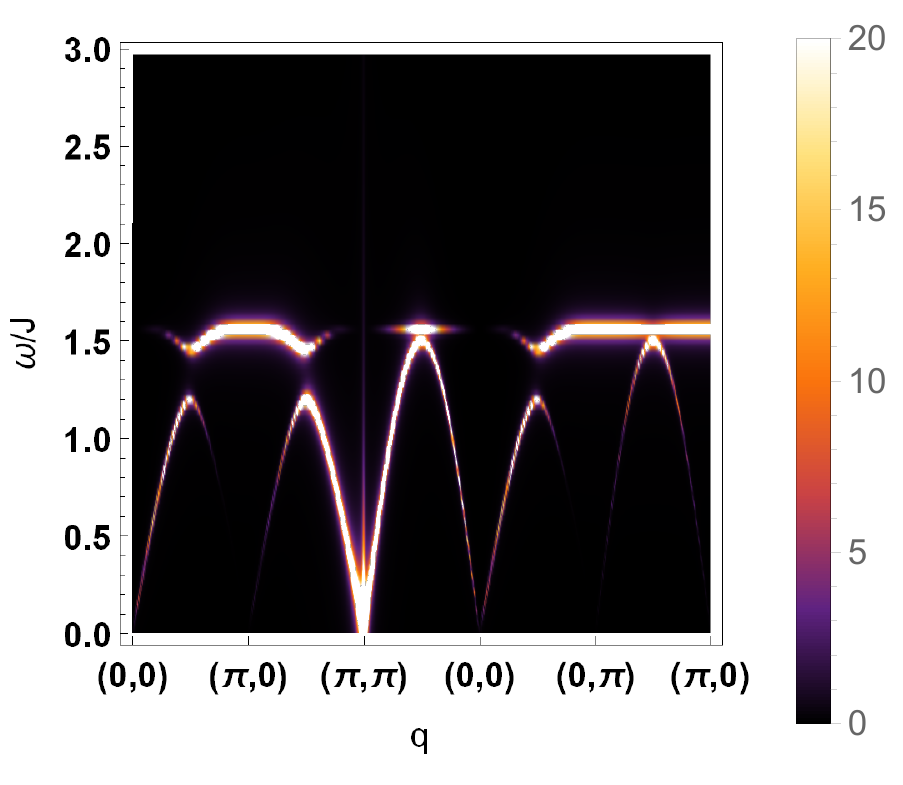}}}
{\subfigure[]{
\includegraphics[width=0.23\textwidth]{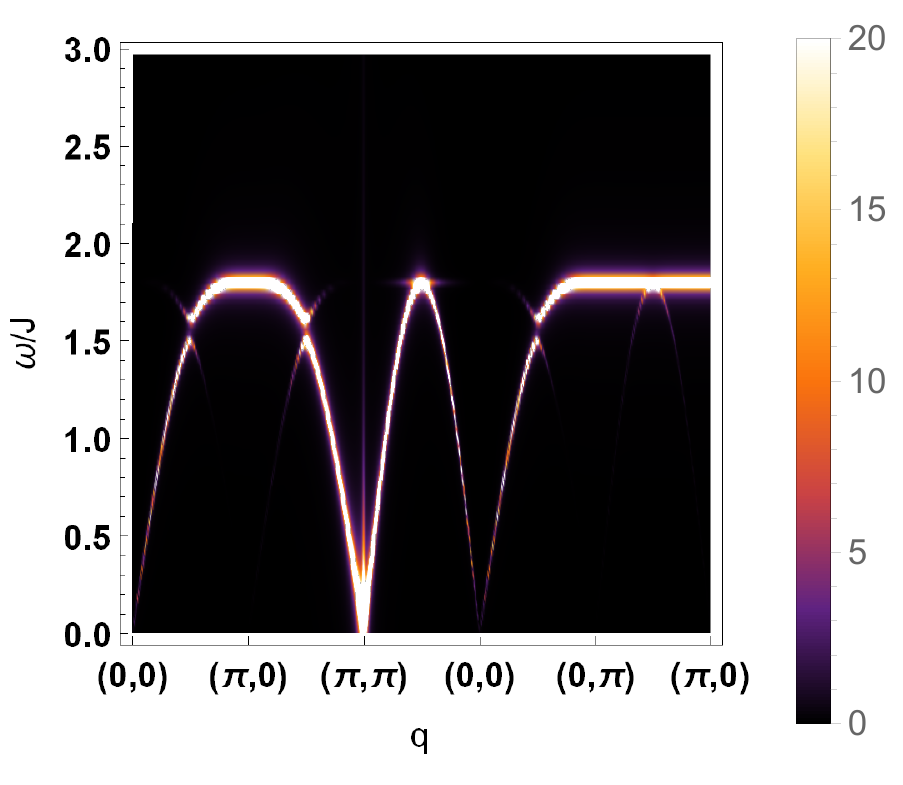}}}
\caption{\label{2-2swt} (Color online) The linear spin wave results of the $2\times 2$ checkerboard model. (a) White dashed lines are the linear spin wave dispersions with $g=0.56$, which are shown here together with the results of Fig.\ref{QMC-SAC/2-2}(b). (b) and (c) are the results of the zero temperature dynamic structure factor with $g=0.56$ and $g=0.8$, respectively, which describe the excitation of a single magnon. Comparing (b) with (c), as $g$ decreases and approaches to $g_c$, the spectral weight of the acoustic branch increases and the low-energy branch becomes visible.}
\end{figure}

In the $2\times 2$ structure, there are two spin wave branches, namely optical and acoustic spin waves, and the magnon dispersions are shown in FIG.\ref{2-2swt}(a). The dispersion of the acoustic branch matches well with the numerical spectrum along $(\pi,0)\to (\pi,\pi)$. The features of the "emergent" branch from $(\pi,0)$ to $(\pi,\pi)$ and the DSF curve along $(\pi,0) \rightarrow (\pi,\pi) \rightarrow (0,0)$ in FIG.\ref{QMC-SAC/2-2}(b) are both captured by LSWT.

It is worthy to know how the spectra evolve when tuning $g$ from 1 to $g_c$, or in other words, when the effect of reduction in the Brillouin zone becomes visible. We find that when $g$ is large, for example $g=0.8$, the LSWT spectra [as shown in FIG.\ref{2-2swt}(c)] are very similar to the bipartite case. The acoustic branch due to an enlarged magnetic unit cell has small spectral weight when deviating from dispersion of the antiferromagnetic square lattice; as $g$ decreases and approaches $g_c$, this low-energy branch becomes visible.

For the spectra of the $2\times 3$ and $2\times 4$ structures, there are many branches of spin waves since their magnetic unit cells contain many spins. For the $2\times 3$ structure, when $\omega<1.5$, an asymmetric DSF curve with a shoulder peak along $(0,0) \rightarrow (\pi,0) \rightarrow (\pi,\pi/3)$ is observed in FIG.\ref{QMC-SAC/3&4}(b), and it is formed by several branches of spin waves as in FIG.\ref{3&4swt}(c). Moreover, for the $2\times 4$ structure, the LSWT result in FIG.\ref{3&4swt}(d) also matches the numerical spectra structure in FIG.\ref{QMC-SAC/3&4}(e) from $(\pi,0)$ to $(\pi,\pi)$ quite well when $\omega<1$.

From the above comparisons, we find that some low-energy structures can be explained qualitatively by LSWT. But comparing the theoretical spectra with corresponding numerics carefully, it is easily found that there are also some mismatches in spectral weight in LSWT. For example, in the $2\times 2$ structure,  LSWT predicts very small spectral weight near $(\pi,0)$ from $(\pi,\pi)$ to $(\pi,0)$ compared to the numerical spectra. These mismatches could be understood as follows. As mentioned above, we find the spectral weight of some low-energy branches become larger when approaching $g_c$ based on spin wave theory. According to the sum rules of the spectrum, it is expected that the spectral weight of some low-energy branches will become much more prominent and some will become smaller in quantum criticality. And it is known that quantum fluctuation becomes very important near quantum criticality, which means the higher-order expansions in spin operator should play an important role in this case. Then the spectral weight, i.e., the matrix element of the spin operator, will be modified a lot compared to the linear spin wave approximation. Finally, interaction terms also change the spectral weight to a certain extent as reported in Refs.\cite{Igarashi,Ma-SAC2018}. However here the goal is to have a qualitative understanding of the low-energy part of the spectra, so we keep our discussion on the linear spin wave level.

\subsubsection{$g< g_c$}
Spectra are gapped when $g<g_c$, which mean that those systems are in the disordered phase. But the overall shapes of spectra do not change a lot compared with the critical spectra except for the enhancement of spectral weight along some paths of momentum; for example the spectral weight is enhanced along $(0,0) \to (\pi,0)$ and $(\pi,0) \to (\pi,\pi/2)$ in FIG.\ref{QMC-SAC/2-2}(c).

To obtain the gapped spectra, one can use spin wave theory and phenomenologically introduce a chemical potential of the bosons and tune the chemical potential to open a gap, or use modified spin wave theory to calculate the gap self-consistently\cite{tang, takahashi, Kogatheory}. But as is known, low-energy excitations in these disordered phases are no longer gapless magnons; they are gapped spin-$1$ triplons which can be described by bond operator theory\cite{sachdev-bond,Zhitomirsky,Uedaprb,Dorettoprb,Kumartriplon}. For simplicity, we consider low-energy excitations in the $2\times 2$ structure (this picture should work for $2\times 3$ and $2\times 4$ also, except for $3\times 3$ which we discuss below separately). They can be understood as follow: when $g=0$ the lattice decouples into isolated plaquettes, and the ground state of such a plaquette is a superposition state consisting of a pair of two-spin singlets along the edges of the plaquette (the explicit expression is given in Appendix.\ref{isola}). The first excited state is formed by breaking a bond; it is a direct product state of a singlet and a triplet. These excitations are localized; once $g$ is switched on they can hop and lower the energy. But the energy required is still finite due to the confinement of spinons\cite{sachdev2008nphys}; this binding energy between a pair of confined spinons leads to the gap in the spectra.

\subsection{High-energy continuum\label{HEC}}
From above comparisons, LSWT describes the low-energy part of the spectra quite well. But this does not mean that the complete spectra of checkerboard lattices behave as single particle with slight modifications. However, LSWT can be improved by considering $1/S$ expansion and performing high-order perturbation to obtain a multi-particle continuum\cite{canali,Igarashi,Lorenzana}, such as a 3-magnon continuum in transverse DSF and a 2-magnon continuum in longitudinal DSF, but the results are not satisfactory even in the bipartite case. This standard perturbation method converges slowly\cite{thirdorder}, and cannot give a reasonable prediction for the energy difference between $(\pi,0)$ and $(\pi/2,\pi/2)$ and spectral weight at these momenta in the bipartite case. Many theories have been proposed to account for these anomalies, for example nearly deconfined spinons \cite{Shao-SAC2017,LJXspinon}, non-perturbative renormalization of magnons \cite{powalski}, singlons \cite{Ayu-singlon}. These mechanisms are quite different but they do not disprove each other and the nature of the high-energy excitation is still in debate. In this section, we study the high-energy part of the spectra in the checkerboard model.

 \begin{figure}
\centering
\includegraphics[width=0.3\textwidth]{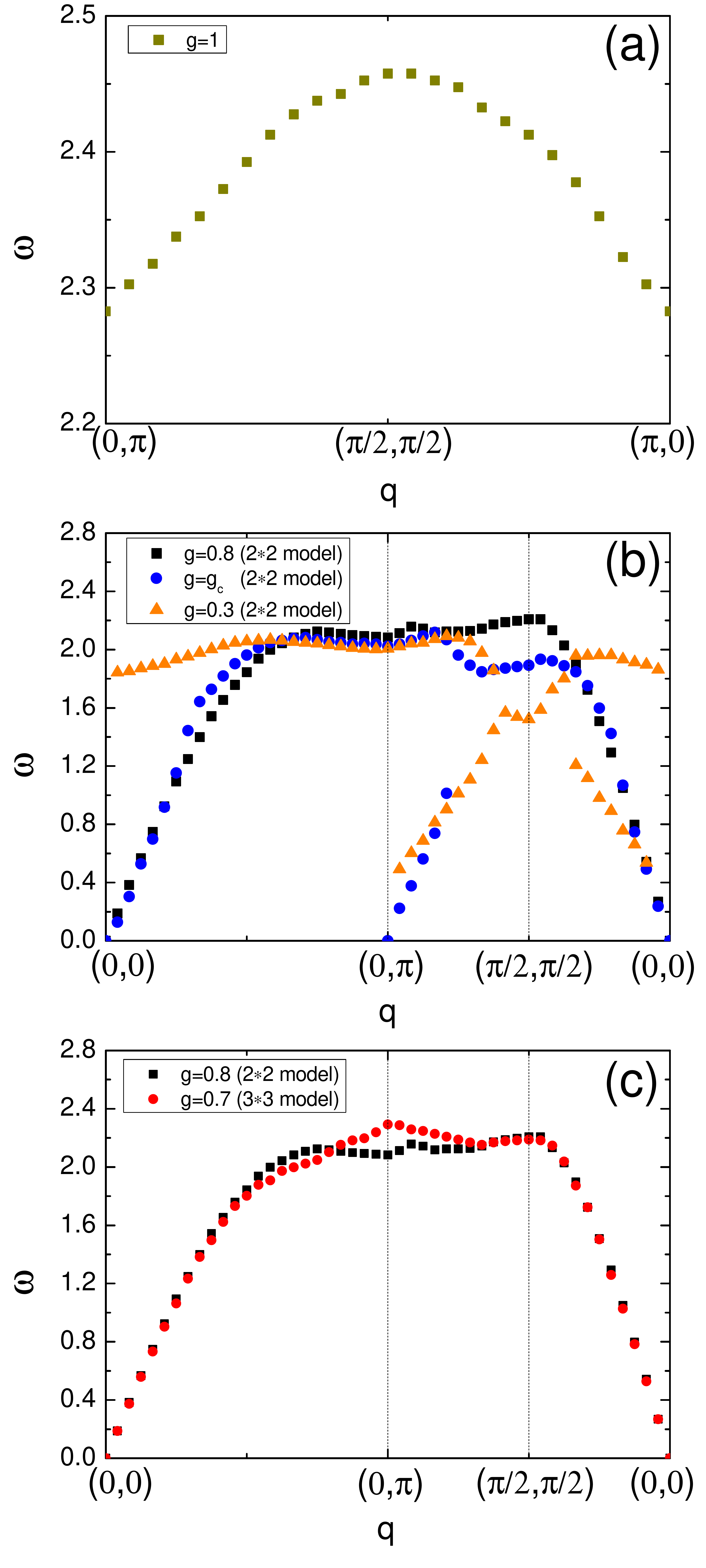}
\caption{\label{disp/2&3} (Color online) The dispersion curve extracted from the QMC-SAC numerical results. (a) All the checkerboard models here reduce to the original square lattice antiferromagnetic Heisenberg model when $g=1$, and the dispersion of the $(\pi/2,\pi/2) \rightarrow (\pi,0)$ is widely concerned in this case. (b) The dispersion curve of $2\times 2$ checkerboard model with varying $g$, where the high-energy dispersion curve shows a larger dip at $(\pi/2,\pi/2)$ as $g$ decreases. (c) A comparison between the dispersion curves of the $2\times 2$ and $3\times 3$ checkerboard models in the N$\mathrm{\acute{e}}$el phase. When the structure of the model changes, the excitation of these two models at $(0,\pi )$ is different.}
\end{figure}

\begin{figure}
\centering
{\subfigure[]{
\includegraphics[scale=0.38]{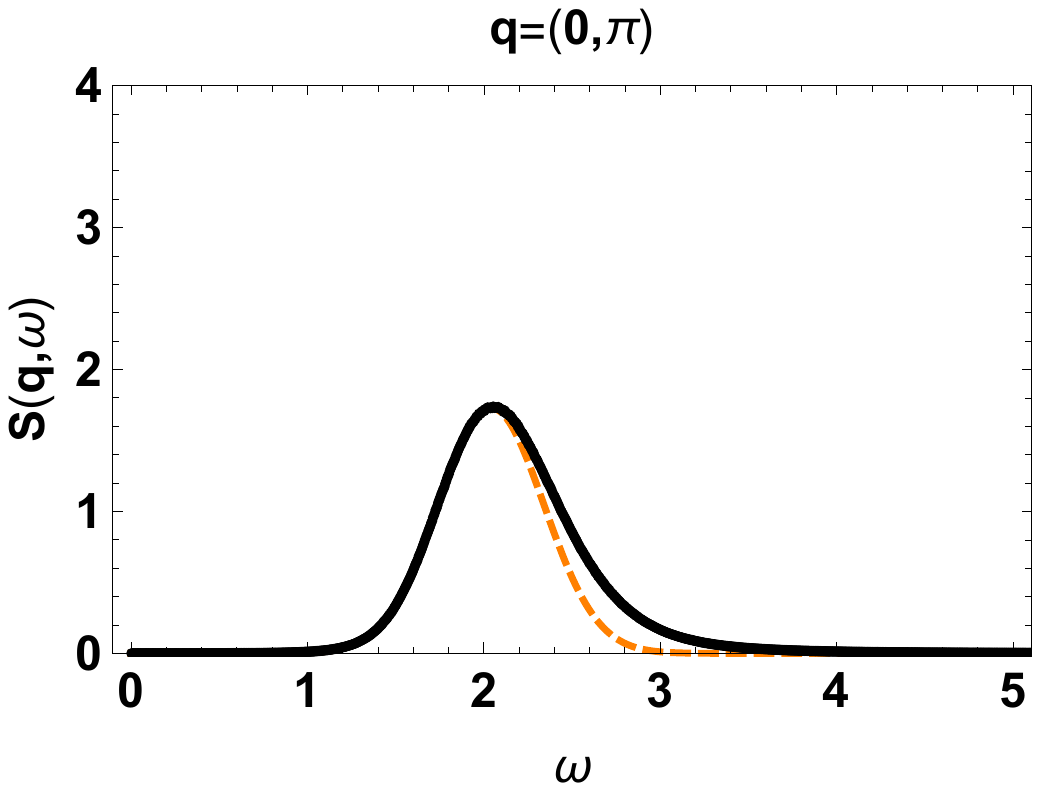}}}
{\subfigure[]{
\includegraphics[scale=0.38]{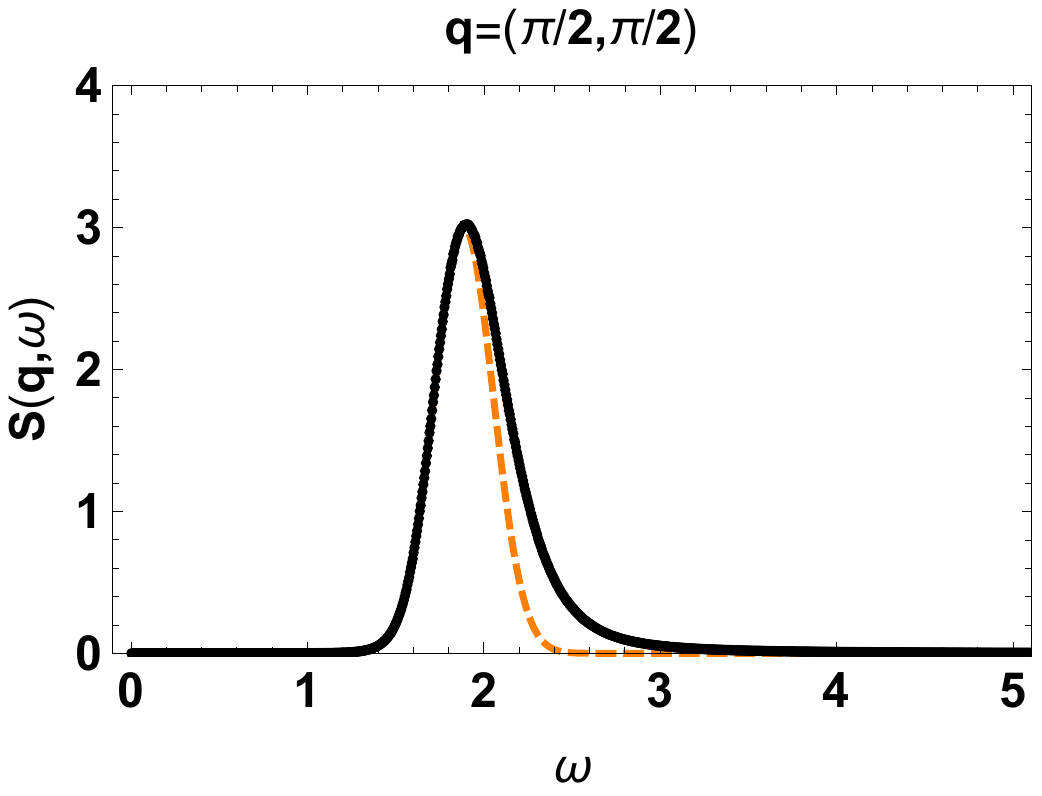}}}
{\subfigure[]{
\includegraphics[scale=0.38]{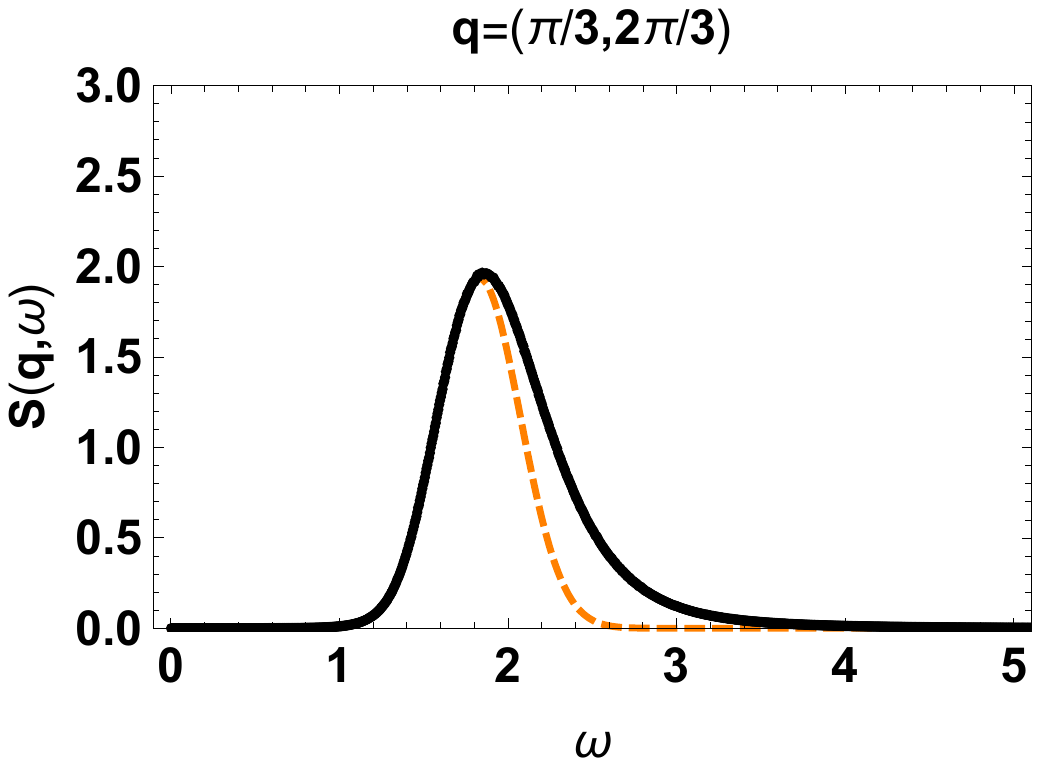}}}
{\subfigure[]{
\includegraphics[scale=0.38]{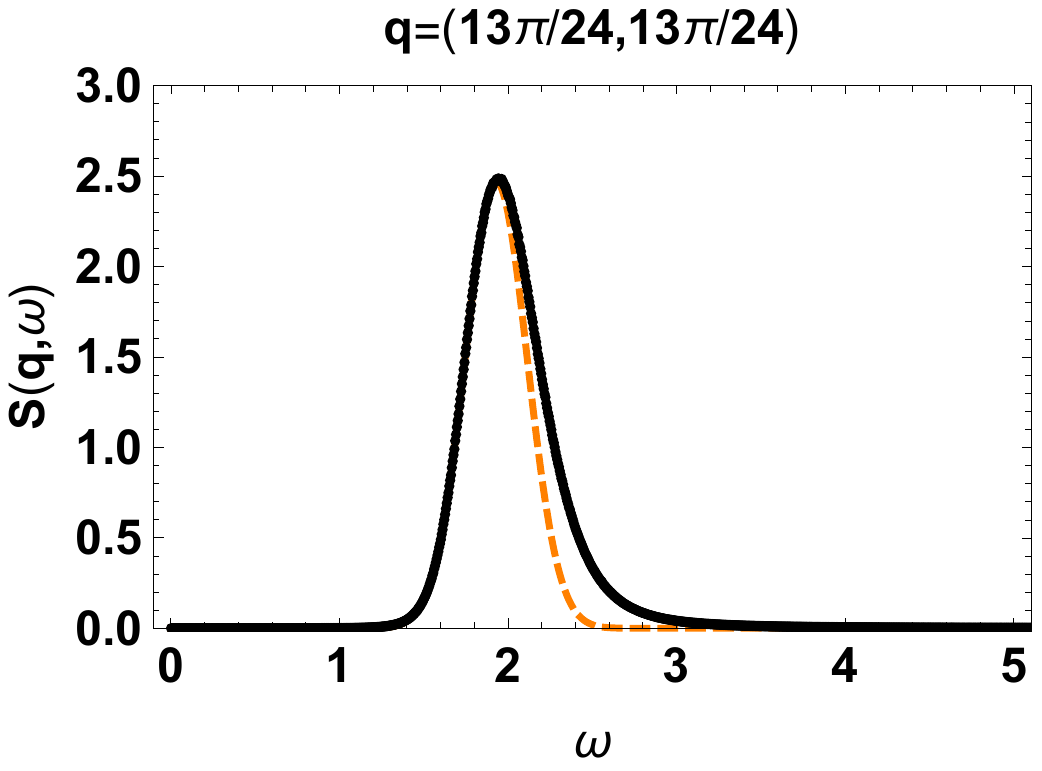}}}
\caption{\label{2-2spec} (Color online) $S(q,\omega)$ of the $2\times 2$ checkerboard model with $g=g_c$, where $q$ is taken as $(0,\pi)$ (a), $(\pi/2,\pi/2)$ (b), $(\pi/3,2\pi/3)$ (c), and $(13\pi/24,13\pi/24)$ (d), which are the special points observed in Fig.\ref{disp/2&3}(b). The black curve is obtained from the QMC-SAC method. The orange dashed curve is Gaussian fitting which can be regarded as a single magnon peak. If the Gaussian peak of the single magnon excitation is subtracted from the numerically calculated results, the remaining part can represent the spectral weight of other modes of spin excitation.}
\end{figure}

We extract the energy dispersion curves from the numerical data by collecting the energy of the local maximum in spectral weight along the selected momentum path shown above, but the energy extracted in this way will be slightly larger than the correct value owing to the finite width of the “$\delta$ function”. In spite of this, the behavior of the spectra should be correct up to a shift in energy.  All of these lattice reduce to antiferromagnetic square lattices when $g=1$, see FIG.\ref{disp/2&3} (a), the roton-like dip at $(\pi,0)$ is observed, and the energy difference between $(0,\pi)$ and $(\pi/2,\pi/2)$ is estimated to be $6.5\%$, which has a good agreement with experiments in $\mathrm{Cu(DCOO)_{2}\cdot 4D_{2}O}$ \cite{ronnow, Dalla}. But it is should be noticed that these extracted energies are slightly larger than the experimental values as expected; the experimental values are $\omega_{(\pi,0)}/J=2.19$ and $\omega_{(\pi/2,\pi/2)}/J=2.38$.

First, we study the effect of structure on the anomaly in the magnetic Brillouin zone (MBZ) boundary. When $g$ is large, spectra of all structures seem like the bipartite case. The remarkable continua in each structure [Fig.\ref{QMC-SAC/2-2} (a), Figs.\ref{QMC-SAC/3&4} (a) and 3(d), FIG.\ref{QMC-SAC/3-3} (a)] are around $(\pi,0)$ and $(0,\pi)$, and they extend from $\omega \approx 1.8$ up to $\omega \approx 3.5$, which is close to the phenomenon observed in square lattices \cite{Dalla}. When $g=g_c$, the low-energy branches appear, which means that the effect of reduction in the MBZ becomes evident; we expect that the continuum should also appear around  endpoints of the magnetic Brillouin zone boundary, for instance, $(\pi/2,\pi/2)$ in $2\times 2$ and $2\times 3$. But the numerical results deviate significantly from our expectation; the most prominent continua are still around $(\pi,0)$ and $(0,\pi)$ [FIig.\ref{QMC-SAC/2-2}(b), Figs.\ref{QMC-SAC/3&4}(b) and (e), FIG.\ref{QMC-SAC/3-3}(b)]. This may relate to the nearly deconfined mechanism. To have a further look, we compare the extracted "dispersion" along $(0,0) \to (0,\pi) \to (\pi/2,\pi/2) \to (0,0)$ for the $2\times 2$ and $3\times 3$ structures in the ordered phase; i.e. in the large-$g$ case, and they are shown in FIG.\ref{disp/2&3}(c). It should be noticed that they are not the true dispersion; the magnetic unit cell is enlarged when $g<1$, so there should be many more branches of dispersion, although they can hardly be extracted from the spectra completely. Despite this, we know that the high-energy optical branches are nearly flat from spin wave calculation; thus it will make sense to study the high-energy part of the "dispersion". There are several extrema in the high-energy part of the “dispersion”, and the anomaly at $(0,\pi)$ is found to be structure-dependent. The energy at $(\pi,0)$ is enhanced in the $3\times 3$ structure compared with $(\pi/2,\pi/2)$, while it is reduced in the $2\times 2$ structure.

Due to the similarity among these spectra of different checkerboard structures, we discuss the effect of $g$ in the $2\times 2$ structure for simplicity. Dispersions with different $g$ are shown in FIG.\ref{disp/2&3} (b), where the dip at $(0,\pi)$ persists in each phase and does not change so much. It is noteworthy that the variation in energy is much more prominent as $g$ changes at $(\pi/2,\pi/2)$ than $(\pi,0)$. At $g=g_c$, the $S(q,\omega)$ at several momenta are shown in FIG.\ref{2-2spec}. They all behave as asymmetric peak, the main contribution can be ascribed to a single optical magnon peak, and the remained part is from the high-energy continuum. As mentioned in Sect. \ref{numerical}, the width of the continua become narrower as $g$ decreases, so the high-energy tail is suppressed.

Finally, we provide a potential account of the high-energy continuum in the checkerboard model investigated here with $g\geq g_c$. Because of the complexity of these checkerboard structures, there are many mechanisms that may lead to these continua. (A) Energy are close for those optical modes, which means there are strong renormalizations among these modes when considering interactions; then they span a larger high-energy range for multimagnon continua. (B) Pairs of spin tend to form singlets due to the checkerboard structure, especially when $g$ is small; then gapped excitation from these singlets may contribute to the high-energy part also\cite{Ayu-singlon}. (C) Nearly deconfined spinons \cite{Shao-SAC2017} may also exist in these models intrinsically. Then a question can be raised: are these spinons deconfined from different optical magnon modes are the same, or are there different modes of spinons? So these lattices provide a playground for studying high-energy continuum, and they also have a close relation to the original square lattice. An open question is how to identify the intrinsic properties of excitations in these high-energy continua.

\subsection{$3\times 3$ model} \label{3-3model}
As mentioned above, the $3\times 3$ model is different from the others. First, no phase transition is observed based on finite-size scaling; i.e., we can not find any finite $g_c$ because of the persistent long-range order. Second, the ground state of an isolated sublattice is a doublet rather than singlet. From these perspectives, the spectrum is very unusual in this structure when $g$ is small; it consists of a low-energy gapless branch and a gapped high-energy part as shown in FIG.\ref{QMC-SAC/3-3}(c).

\begin{figure}
\centering
\includegraphics[scale=0.8]{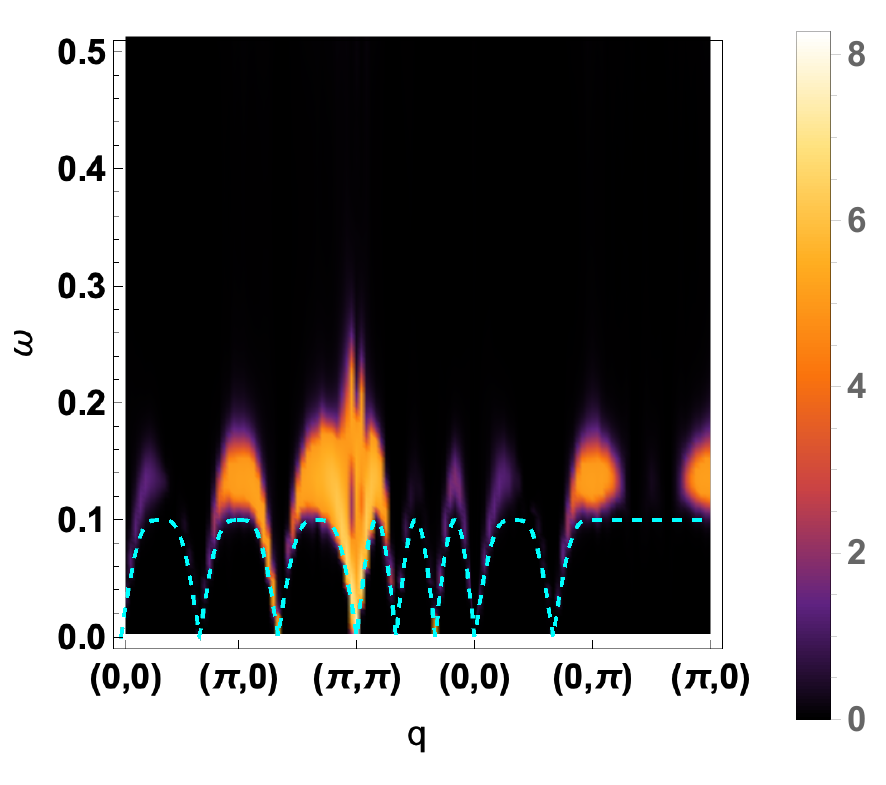}
\caption{\label{3-3lsw} (Color online) Low-energy spectra extracted from Fig.\ref{QMC-SAC/3-3}.(c), i.e., the $3\times 3$ checkerboard model with $g=0.1$. The cyan dashed line is calculated by linear spin wave theory. A noteworthy feature is the prominent continuum centered at the endpoints of the magnetic Brillouin zone boundary, for example, at $(\pi,0)$ and $(\pi,2\pi/3)$.}
\end{figure}

For a better understanding of the excitation in this case, the spectrum of a few $3\times 3$ plaquettes is the key. But it is hard to write down the explicit wave functions for the ground state and low-lying excited states even for such an isolated plaquette, so we adopt an exact diagonalization; the energy levels are given in Appendix \ref{isola}. According to the exact diagonalization result in FIG.\ref{energylevel}(a), the ground state, the first excited state, and the second excited state of the isolated $3\times 3$ plaquette are an $S=1/2$ doublet, $S=3/2$ quartet, and  another $S=1/2$ doublet, respectively. The gap between the ground state and first excited state is quite large, and it equals $J_1$. Let us turn to the small-$g$ case, more precisely, $g\ll 1$; then $g$ can be considered as a perturbation in this circumstance. Once $g$ is turned on, the original doublet ground state in the isolated case will split into some lower-energy states and some higher-energy states. For example, let us consider two $3\times 3$ plaquettes. The ground state of this system is a singlet, and the first excited state is a triplet; they are formed by the original $S=1/2$ ground state of each plaquette, and the energy difference between the ground state and first excited state is around $g$. The following excited states are formed by combining an original $S=1/2$ ground state and an original $S=3/2$ first excited state, so they should be $S=1$ and $S=2$ states with energy around $J_1$. This picture is proved by exact diagonalization, see FIG.\ref{energylevel}(b). For $2\times 2$, $2\times 3$ and $2\times 4$ models, the ground states of the isolated plaquette are singlet, so their ground states will not split even $g$ is turned on. One more interesting thing is that: in small $g$ case, the ground state of $3\times 3$ model depends on the size \cite{lieb-mattis};  the ground state is singlet or doublet depending on even or odd size, but this fact will not change the low-lying excitation.

As discussed above, the low-lying excited states splitting from original doublet are separated from the higher-energy states by a gap with energy around $J_1$, and $\Delta S$ between the ground state and first excited state should be 1. For this reason, we considered the $S=1$ excitation from the ground state to first excited state first. As a result of the existence of the low-energy gapless branch and $S=1$ excitation, we find that this gapless branch can be fitted by a spin wave with an enlarged lattice constant very well without other parameters; specifically, the dispersion is $g\sqrt{1-\gamma_k^{2}}$, where $\gamma_{k}=(\cos{3k_x}+\cos{3k_y})/2$ (see FIG.\ref{3-3lsw}). So this $S=1$ excitation can be regarded as a spin wave in N$\mathrm{\acute{e}}$el order formed by a “block spin” in each $3\times 3$ sublattice with an effective exchange interaction originating from renormalization. The effective exchange interaction between these “block spin” is not as simple as only a nearest-neighbor interaction; next-nearest-neighbor or even ring-exchange interactions may exist also. What surprised us also is that the continuum centered at the endpoints of the magnetic Brillouin zone boundary are still prominent, for example, at $(\pi,0)$ and $(\pi,2\pi/3)$. Then, a question is, what mechanism leads to this low-energy continuum? Does the nearly deconfined account still work for such renormalized “block spin”? What is the contribution to the continuum from the “internal” structure of this “block spin”?

For the remaining part, namely gapped excitation, compared to other structures, the low-energy features such as the periodic structure and especially the shape around $(\pi,\pi)$ in the ordered phase are almost lost when $g$ is small. Based on the exact diagonalization result, this gapped continuum is from mixing multiplet state excitations, and they should be excitations mainly concentrated in the sublattice as discussed in Sec. \ref{numerical}, so it is hard to find well-defined dispersion behavior from the low-energy part of this continuum. It is noteworthy that we find such a system with the coexistence of magnon and higher multiplet excitations and even nearly deconfined spinons, which deserves further study for complete understanding of the anomalous high-energy continuum and nearly deconfined mechanism \cite{groundstate}.

\section{Conclusions}\label{conclusion}
In this work, we have investigated the dynamic spin structure factor of the spin-1/2 antiferromagnetic Heisenberg model in several checkerboard structures, from the ordered phase to disordered phase as well as at the critical point, by using the QMC-SAC method. The models we studied can be sorted into two classes according to whether there is a phase transition at finite $g_c$ or not; $2\times 2$, $2\times 3$, and $2\times 4$ belong to the same class, while $3\times 3$ is in another class.

For both classes of checkerboard models, when $g$ is large, the spectra of all structures behave as an antiferromagnetic square lattice with a prominent high-energy continuum around $(\pi,0)$ and $(0,\pi)$. Due to the enlarged unit cell, the high-energy continuum can be contributed by different excitations, such as optical magnons and nearly deconfined spinons; some improved theory is needed to have a better understanding.

For models with a finite $g_c$, when approaching $g_c$ from the ordered phase, the low-energy branches due to enlarged magnetic unit cells become visible, and they can be described by liner spin wave theory quite well. When $g<g_c$, their spectra are gapped, but the overall shapes do not change a lot compared with corresponding critical spectra except for enhancement of spectral weight around some momenta.

There is no phase transition in the $3\times 3$ checkerboard model owing to the persistent long-range order.  The doublet ground state of an isolated $3\times 3$ plaquette leads to unusual spectra. A gap between the gapless branch and high-energy part exists in this structure when $g$ is small. The high-energy continuum consists of mixed multiplet excitations. The gapless branch in this case can be regarded as a spin wave in N$\mathrm{\acute{e}}$el order formed by the “block spin” in each $3\times 3$ plaquette with effective exchange interaction originating from renormalization. The effective exchange interaction between these “block spin” is not as simple as only nearest-neighbor interaction; next-nearest-neighbor or even ring-exchange interaction may exist also. One noteworthy finding is that the continuum also appears in this low-energy branch.

Finally, a $2\times 2$ plaquette-like lattice and $3\times 3$ cluster have been realized in experiments; we would like to compare our theoretical results with further experimental spectra to gain insight into the understanding of these complicated excitations.

\begin{acknowledgements}
We thank Yi-Zhuang You, Anders W. Sandvik, Shuai Yin, Peng Ye, Guang-Ming Zhang, and Trinanjan Datta for helpful discussions.  Y. X, Z. X, and D.X.Y. are supported by the National Key R$\&$D Program of China 2017YFA0206203, 2018YFA0306001, NSFC-11574404, NSFC-11275279, Three Big Constructions-Supercomputing Application Cultivation Projects, and the Leading Talent Program of Guangdong Special Projects. H. Q. Wu is supported by NSFC-11804401 and the start-up funding of SYSU.

\end{acknowledgements}

\appendix
\section{Linear spin wave theory} \label{appendix}
 There are many bosonization transformations for spin operators. Dyson-Maleev and Holstein-Primakoff transformation are widely used, they are the same at the linear spin wave level. The spin operators at the linear spin wave level are expressed in terms of boson operators as
\begin{equation}
\begin{aligned}\label{bosonization}
S_{i}^{z}=&S-a_{i}^{\dagger}a_{i},  \, S_{i}^{+}\approx \sqrt{2S}a_{i},  \, S_{i}^{-}\approx \sqrt{2S}a_{i}^{\dagger}, \\
S_{j}^{z}=&b_{j}^{\dagger}b_{j}-S,  \, S_{j}^{+}\approx \sqrt{2S}b_{j}^{\dagger}, \, S_{j}^{-}\approx \sqrt{2S}b_{j},
\end{aligned}
\end{equation}
where $a_{i}^{\dagger}$ , $a_{i}$ are for up spin, $b_{j}^{\dagger}$ , $b_{j}$ for down spin. The linear spin wave Hamiltonian $H_{2}$ ($H_{2}$ means quadratic form) can be obtained by using Eq.(\ref{bosonization}) to express the original Hamiltonian Eq.(\ref{Hamiltonian}) in terms of boson operators, then introducing the Fourier transformation of the boson operator. $H_{2}$ of the checkerboard lattices investigated in this paper are given as follows:

(1). $2\times 2$:
\begin{equation}
\begin{aligned}\label{2by2ham}
H_{2}=&SJ\sum_{\textbf{k}}\{2(1+g)(a_{\textbf{k}}^{+}a_{\textbf{k}}+b_{-\textbf{k}}^{+}b_{-\textbf{k}}+c_{-\textbf{k}}^{+}c_{-\textbf{k}}
+d_{\textbf{k}}^{+}d_{\textbf{k}})\\
&+[\gamma(k_x)(a_{\textbf{k}}b_{-\textbf{k}}+d_{\textbf{k}}^{+}c_{-\textbf{k}}^{+})
+\gamma(k_y)(a_{\textbf{k}}c_{-\textbf{k}}+d_{\textbf{k}}^{+}b_{-\textbf{k}}^{+})\\
&+h.c]\},
\end{aligned}
\end{equation}
where $a_{\textbf{k}},b_{\textbf{k}},...$ are boson operators for A,B,... sites as depicted in FIG.\ref{model-structure}(a),
$\gamma(k)=e^{-i\, k}+g\,e^{i\, k}$, and h.c. means Hermitian conjugate. Then using the standard method\cite{Hemmen} to diagonalize the bosonic quadratic Hamiltonian Eq.(\ref{2by2ham}) to obtain the linear spin wave dispersion, we have checked that our results are equivalent to Koga's results \cite{Kogatheory} .

(2). $2 \times 3$:
\begin{equation}
\begin{aligned}
H_{2}=&SJ\sum_{\textbf{k}}\{2(1+g)(a_{\textbf{k}}^{+}a_{\textbf{k}}+b_{-\textbf{k}}^{+}b_{-\textbf{k}}+g_{-\textbf{k}}^{+}g_{-\textbf{k}}
+l_{\textbf{k}}^{+}l_{\textbf{k}}\\
&+h_{\textbf{k}}^{+}h_{\textbf{k}}+k_{-\textbf{k}}^{+}k_{-\textbf{k}}+e_{\textbf{k}}^{+}e_{\textbf{k}}
+f_{-\textbf{k}}^{+}f_{-\textbf{k}})+(3+g)(c_{-\textbf{k}}^{+}c_{-\textbf{k}}\\
&+d_{\textbf{k}}^{+}d_{\textbf{k}}+i_{\textbf{k}}^{+}i_{\textbf{k}}+j_{-\textbf{k}}^{+}j_{-\textbf{k}})+[\gamma(k_x)(a_{\textbf{k}}b_{-\textbf{k}}+d_{\textbf{k}}^{+}c_{-\textbf{k}}^{+}\\
&+e_{\textbf{k}}f_{-\textbf{k}}+h_{\textbf{k}}^{+}g_{-\textbf{k}}^{+}+i_{\textbf{k}}j_{-\textbf{k}}+l_{\textbf{k}}^{+}k_{-\textbf{k}}^{+})+\gamma_{1}(k_y)(a_{\textbf{k}}c_{-\textbf{k}}\\
&+d_{\textbf{k}}f_{-\textbf{k}}+d_{\textbf{k}}^{+}b_{-\textbf{k}}^{+}+e_{\textbf{k}}^{+}c_{-\textbf{k}}^{+}
+g\,e_{\textbf{k}}g_{-\textbf{k}}+g\,h_{\textbf{k}}^{+}f_{-\textbf{k}}^{+}\\
&+i_{\textbf{k}}^{+}g_{-\textbf{k}}^{+}+h_{\textbf{k}}j_{-\textbf{k}}+
i_{\textbf{k}}k_{-\textbf{k}}+l_{\textbf{k}}^{+}j_{-\textbf{k}}^{+}+g\,a_{\textbf{k}}^{+}k_{-\textbf{k}}^{+}+g\,l_{\textbf{k}}b_{-\textbf{k}})\\
&+h.c.]\},
\end{aligned}
\end{equation}
here, $\gamma(k_x)=e^{-i\,k_x}+g\,e^{i\,k_x}$ and $\gamma_{1}(k_y)=e^{-i\,k_y}$, there are 12 sites in a magnetic unit cell in this structure .

(3). $2\times 4$:
\begin{equation}
\begin{aligned}
H_{2}=&SJ\sum_{\textbf{k}}\{2(1+g)(a_{\textbf{k}}^{+}a_{\textbf{k}}+b_{-\textbf{k}}^{+}b_{-\textbf{k}}+g_{-\textbf{k}}^{+}g_{-\textbf{k}}
+h_{\textbf{k}}^{+}h_{\textbf{k}})\\
&+(3+g)(c_{-\textbf{k}}^{+}c_{-\textbf{k}}+d_{\textbf{k}}^{+}d_{\textbf{k}}+e_{\textbf{k}}^{+}e_{\textbf{k}}+
f_{-\textbf{k}}^{+}f_{\textbf{k}})+[\gamma(k_x)\\
&(a_{\textbf{k}}b_{-\textbf{k}}+e_{\textbf{k}}f_{-\textbf{k}}+d_{\textbf{k}}^{+}c_{-\textbf{k}}^{+}
+h_{\textbf{k}}^{+}g_{-\textbf{k}}^{+})
+\gamma_{1}(k_y)(a_{\textbf{k}}c_{-\textbf{k}}+d_{\textbf{k}}f_{-\textbf{k}}\\
&+e_{\textbf{k}}g_{-\textbf{k}}
+g\,h_{\textbf{k}}b_{-\textbf{k}}+e_{\textbf{k}}^{+}c_{-\textbf{k}}^{+}+h_{\textbf{k}}^{+}f_{-\textbf{k}}^{+}+d_{\textbf{k}}^{+}b_{-\textbf{k}}^{+}
+g\,a_{\textbf{k}}^{+}g_{-\textbf{k}}^{+})\\
&+h.c.]\},
\end{aligned}
\end{equation}
with $\gamma(k_x)=e^{-i\,k_x}+g\,e^{i\,k_x}$ and $\gamma_{1}(k_y)=e^{-i\,k_y}$.

(4). $3\times 3$:
\begin{equation}
\begin{aligned}
H_{2}=&SJ\sum_{\textbf{k}}\{2(1+g)(a_{\textbf{k}}^{+}a_{\textbf{k}}+g_{\textbf{k}}^{+}g_{\textbf{k}}+c_{\textbf{k}}^{+}c_{\textbf{k}}+i_{\textbf{k}}^{+}i_{\textbf{k}}+j_{-\textbf{k}}^{+}j_{-\textbf{k}}\\
&+p_{-\textbf{k}}^{+}p_{-\textbf{k}}+r_{-\textbf{k}}^{+}r_{-\textbf{k}}+l_{-\textbf{k}}^{+}l_{-\textbf{k}})+(3+g)(b_{-\textbf{k}}^{+}b_{-\textbf{k}}\\
&+d_{-\textbf{k}}d_{-\textbf{k}}+f_{-\textbf{k}}^{+}f_{-\textbf{k}}+h_{-\textbf{k}}^{+}h_{-\textbf{k}}+k_{\textbf{k}}^{+}k_{\textbf{k}}+m_{\textbf{k}}^{+}m_{\textbf{k}}+o_{\textbf{k}}^{+}o_{\textbf{k}}\\
&+q_{\textbf{k}}^{+}q_{\textbf{k}})+4(e_{\textbf{k}}^{+}e_{\textbf{k}}+n_{-\textbf{k}}^{+}n_{-\textbf{k}})+[\gamma_{1}(k_x)(a_{\textbf{k}}b_{-\textbf{k}}+c_{\textbf{k}}^{+}b_{-\textbf{k}}^{+}\\
&+e_{\textbf{k}}^{+}d_{-\textbf{k}}^{+}+e_{\textbf{k}}f_{-\textbf{k}}+g_{\textbf{k}}h_{-\textbf{k}}+i_{\textbf{k}}^{+}h_{-\textbf{k}}^{+}+k_{\textbf{k}}^{+}j_{-\textbf{k}}^{+}+k_{\textbf{k}}l_{-\textbf{k}}\\
&+m_{\textbf{k}}n_{-\textbf{k}}+o_{\textbf{k}}^{+}n_{-\textbf{k}}^{+}+q_{\textbf{k}}^{+}p_{-\textbf{k}}^{+}+q_{\textbf{k}}r_{-\textbf{k}}+g a_{\textbf{k}}^{+}l_{-\textbf{k}}^{+}+g o_{\textbf{k}}d_{-\textbf{k}}\\
&+g g_{\textbf{k}}^{+}r_{-\textbf{k}}^{+}+g c_{\textbf{k}}j_{-\textbf{k}}+g m_{\textbf{k}}^{+}f_{-\textbf{k}}^{+}+g i_{\textbf{k}}p_{-\textbf{k}})+\gamma_{1}(k_y)(a_{\textbf{k}}d_{-\textbf{k}}\\
&+e_{\textbf{k}}^{+}b_{-\textbf{k}}^{+}+c_{\textbf{k}}f_{-\textbf{k}}+g_{\textbf{k}}^{+}d_{-\textbf{k}}^{+}+e_{\textbf{k}}h_{-\textbf{k}}+i_{\textbf{k}}^{+}f_{-\textbf{k}}^{+}+m_{\textbf{k}}^{+}j_{-\textbf{k}}^{+}\\
&+k_{\textbf{k}}n_{-\textbf{k}}+o_{\textbf{k}}^{+}l_{-\textbf{k}}^{+}+m_{\textbf{k}}p_{-\textbf{k}}+q_{\textbf{k}}^{+}n_{-\textbf{k}}^{+}+o_{\textbf{k}}r_{-\textbf{k}}+g a_{\textbf{k}}^{+}p_{-\textbf{k}}^{+}\\
&+g q_{\textbf{k}}b_{-\textbf{k}}+g c_{\textbf{k}}^{+}r_{-\textbf{k}}^{+}+g g_{\textbf{k}}j_{-\textbf{k}}+g k_{\textbf{k}}^{+}h_{-\textbf{k}}^{+}+g i_{\textbf{k}}l_{-\textbf{k}})+h.c.]\},
\end{aligned}
\end{equation}
where $\gamma_{1}(k)=e^{-i\,k}$; there are 18 sites in a magnetic unit cell in this structure.

The dynamic structure factor is defined by
\begin{equation}
S^{\alpha \beta}(\textbf{q},\omega)=\int_{-\infty}^{\infty}\frac{dt}{2\pi}\langle S^{\alpha}_{q}(t)S^{\beta}_{-q}(0)\rangle e^{i\,\omega t},
\end{equation}
where $\alpha$, $\beta$ refer to x, y, z. It can be expressed in this form also\cite{Yao-Sq, Shao-SAC2017}:
$S^{\alpha \alpha}(\textbf{q},\omega)=\sum_{f}|\langle f|S^{\alpha}(\textbf{q})|0\rangle|^{2}\delta(\omega-(\omega_{f}-\omega_{0}))$.

\section{isolated plaquette} \label{isola}
\begin{figure}
\centering
\includegraphics[scale=1]{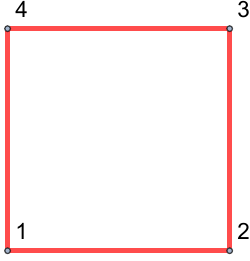}
\caption{\label{isolated} (Color online) An isolated $2\times 2$ plaquette.}
\end{figure}

(1) $2\times 2$ plaquette:
In the $g=0$ limit, the lattice decomposes into an isolated plaquette without inter-plaquette interaction. Consider an isolated plaquette (FIG.\ref{isolated}); the Hamiltonian of this block is given by
\begin{equation}
H_p=J(\textbf{S}_{1}\cdot \textbf{S}_{2}+\textbf{S}_{2}\cdot \textbf{S}_{3}+\textbf{S}_{3}\cdot \textbf{S}_{4}+\textbf{S}_{1} \cdot \textbf{S}_{4}),
\end{equation}
it can be easily solved\cite{Kumartriplon,Uedaprb,Dorettoprb} by introducing: $\textbf{S}_{a}=\textbf{S}_{1}+\textbf{S}_{3}$ and $\textbf{S}_{b}=\textbf{S}_{2}+\textbf{S}_{4}$, $\textbf{S}=\textbf{S}_{a}+\textbf{S}_{b}$ then the Hamiltonian can be written as
\begin{equation}
H_p=\frac{1}{2}J(\textbf{S}^2-\textbf{S}_{a}^{2}-\textbf{S}_{b}^{2}),
\end{equation}
the eigenvalues are given in Table.\ref{tab:isolated22pla}; the eigenstates can be labeled by quantum number $S,M,S_{a},S_{b}$, where $M$ corresponds to $S^{z}$. The ground state is a singlet with $S=0$, $M=0$, $S_{a}=1$, $S_{b}=1$, and energy $-2J_{1}$; the wave function is $([1,2]\otimes [4,3]+[1,4]\otimes [2,3])/\sqrt{3}$.

The first excited state is a triplet with $S=1$, $S_{a}=1$, $S_{b}=1$, and energy $-J_{1}$; the wave functions are $|S=1,M=1\rangle \propto |\uparrow_{1}\uparrow_{2}\rangle \otimes [4,3]+ [2,1]\otimes  |\uparrow_{3}\uparrow_{4}\rangle$, $|S=1,M=0\rangle \propto  [1,2]\otimes \{3,4\}+{1,2}\otimes [3,4]$, and $|S=1,M=-1\rangle \propto  [1,2]\otimes |\downarrow_{3}\downarrow_{4}\rangle + |\downarrow_{1}\downarrow_{2}\rangle \otimes [3,4]$, where $[i,j]=( |\uparrow_{i}\downarrow_{j}\rangle- |\downarrow_{i}\uparrow_{j}\rangle)/\sqrt{2}$, $\{i,j\}=( |\uparrow_{i}\downarrow_{j}\rangle+|\downarrow_{i}\uparrow_{j}\rangle)/\sqrt{2}$.

\begin{figure}
\centering
\includegraphics[width=0.4\textwidth]{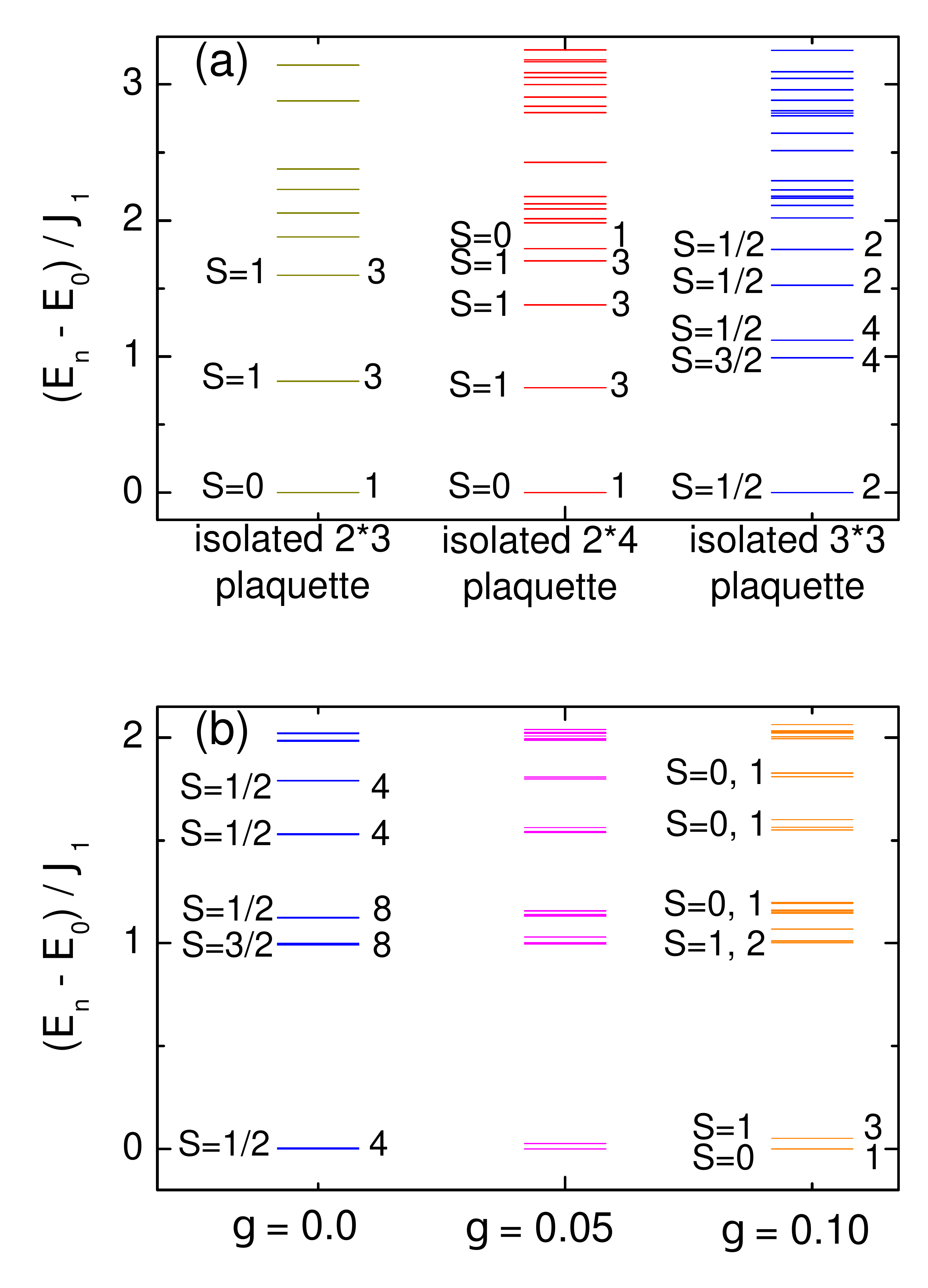}
\caption{\label{energylevel} (Color online) (a) Energy levels of isolated $2\times 3$, $2\times 4$, and $3\times 3$ plaquettes. (b) Energy levels of two $3\times 3$ plaquettes; inter-plaquette interaction is $g$.}
\end{figure}

(2) $2\times 3$, $2\times 4$, and $3\times 3$ plaquettes:

We calculated spectra of these 3 structures by exact diagonalization; the results are shown in FIG.\ref{energylevel}(a). The spectra of two $3\times 3$ plaquettes with different inter-plaquette interactions $g$ are shown in FIG.\ref{energylevel}(b).

\begin{table}
\caption{Eigenvalues of an isolated $2\times 2$ plaquette.}
\begin{ruledtabular}
\label{tab:isolated22pla}
\begin{tabular}{cllll}
$\textbf{S}_{a}$ &  $\textbf{S}_{b}$  & $\textbf{S}$ & Eigenvalues & Degeneracy\\ \hline
0 & 0 & 0 & 0 & 1 \\
0 & 1 & 1 & 0 & 3 \\
1 & 0 & 1 & 0 & 3 \\
1 & 1 & 0 & -2$J$ & 1\\
1 & 1 & 1 & -$J$ & 3 \\
1 & 1 & 2 & $J$ & 5 \\
\end{tabular}
\end{ruledtabular}
\end{table}

\newpage
\bibliographystyle{apsrev4-1}
\bibliography{pla-ref}
\end{document}